\documentclass[twocolumn]{aastex62}

\usepackage{multirow}

\newcommand{\kms}{km s$^{-1}$}
\newcommand{\myr}{$M_{\odot}$ yr$^{-1}$}
\newcommand{\msun}{$M_{\odot}$}
\newcommand{\mstar}{$M_{\star}$}
\newcommand{\mgas}{$M_{\rm gas}$}
\newcommand{\mdust}{$M_{\rm dust}$}
\newcommand{\kkmspc}{K~km~s$^{-1}$~pc$^2$}

\newcommand{\jykms}{Jy~km~s$^{-1}$}
\newcommand{\ci}{[C\,{\footnotesize I}]}
\newcommand{\hi}{H\,{\footnotesize I}}
\newcommand{\co}{CO}
\newcommand{\cione}{[C\,{\footnotesize I}]$(^3P_1\,-\,^{3}P_0)$}
\newcommand{\citwo}{[C\,{\footnotesize I}]$(^3P_2\,-\, ^{3}P_1)$}
\newcommand{\ciplus}{[C\,{\footnotesize II}]}
\newcommand{\cofive}{CO\,$(5-4)$}
\newcommand{\cofour}{CO\,$(4-3)$}
\newcommand{\cothree}{CO\,$(3-2)$}
\newcommand{\cotwo}{CO\,$(2-1)$}
\newcommand{\coone}{CO\,$(1-0)$}
\newcommand{\lprimeci}{$L'_{\mathrm{[C\,\scriptscriptstyle{I}\scriptstyle{]}}^3P_1\,-\, ^3P_0}$}
\newcommand{\lprimecifull}{$L'_{\mathrm{[C\,\scriptscriptstyle{I}\scriptstyle{]}}^3P_1\,-\, ^3P_0}$}
\newcommand{\lprimecitwo}{$L'_{\mathrm{[C\,\scriptscriptstyle{I}\scriptstyle{]}}^3P_2\,-\, ^3P_1}$}
\newcommand{\lprimecotwo}{$L'_{\rm CO(2-1)}$}
\newcommand{\lprimecoone}{$L'_{\rm CO(1-0)}$}
\newcommand{\ici}{$I_{\mathrm{[C\,\scriptscriptstyle{I}\scriptstyle{]}}^3P_1\,-\, ^3P_0}$}
\newcommand{\lir}{$L_{\rm IR}$}
\newcommand{\lfir}{$L_{\rm FIR}$}
\newcommand{\lsun}{$L_{\odot}$}
\newcommand{\mci}{$M_{\rm{[C\,\scriptscriptstyle{I}\scriptstyle{]}}}$}
\newcommand{\fci}{$f_{\rm{[C\,\scriptscriptstyle{I}\scriptstyle{]}}}$}
\newcommand{\ciabundance}{\ci/[H$_2$]}

\received{2018}
\revised{2018}
\submitjournal{ApJ}

\shorttitle{A Survey of Atomic Carbon \textsc{[C\,{\footnotesize I}]} in
  High-redshift Main-Sequence Galaxies}
\shortauthors{Valentino et al.}

\begin{document}

\title{A Survey of Atomic Carbon \textsc{[C\,{\footnotesize I}]} in
  High-redshift Main-Sequence Galaxies}


\correspondingauthor{Francesco Valentino}
\email{francesco.valentino@nbi.ku.dk}

\author[0000-0001-6477-4011]{Francesco Valentino}
\affil{Dawn Cosmic Center, Niels Bohr Institute, University of Copenhagen, Juliane Maries Vej 30, DK-2100 Copenhagen, Denmark}
\affil{Dark Cosmology Centre, Niels Bohr Institute, University of Copenhagen, Juliane Maries Vej 30, DK-2100 Copenhagen, Denmark}

\author[0000-0002-4872-2294]{Georgios E. Magdis}
\affil{Dawn Cosmic Center, Niels Bohr Institute, University of
 Copenhagen, Juliane Maries Vej 30, DK-2100 Copenhagen, Denmark}
\affil{Dark Cosmology Centre, Niels Bohr Institute, University of
 Copenhagen, Juliane Maries Vej 30, DK-2100 Copenhagen, Denmark}
\affil{Institute for Astronomy, Astrophysics, Space Applications and
 Remote Sensing, National Observatory of Athens, GR-15236 Athens, Greece}

\author[0000-0002-3331-9590]{Emanuele Daddi}
\affil{Laboratoire AIM-Paris-Saclay, CEA/DSM-CNRS-Universit\'{e} Paris
 Diderot, Irfu/Service d'Astrophysique, CEA Saclay, Orme des Merisiers, F-91191
 Gif sur Yvette, France}

\author[0000-0001-9773-7479]{Daizhong Liu}
\affil{Max Planck Institute for Astronomy, K\"{o}nigstuhl 17, D-69117 Heidelberg, Germany}

\author[0000-0002-6290-3198]{Manuel Aravena}
\affil{N\'{u}cleo de Astronom\'{i}a, Facultad de Ingenier\'{i}a y
 Ciencias, Universidad Diego Portales, Av. Ejército 441, Santiago, Chile}

\author[0000-0002-5743-0250]{Fr\'{e}d\'{e}ric Bournaud}
\affil{Laboratoire AIM-Paris-Saclay, CEA/DSM-CNRS-Universit\'{e} Paris
 Diderot, Irfu/Service d'Astrophysique, CEA Saclay, Orme des Merisiers, F-91191
 Gif sur Yvette, France}

\author[0000-0003-4578-514X]{Anna Cibinel}
\affil{Astronomy Centre, Department of Physics and Astronomy,
University of Sussex, Brighton, BN1 9QH, UK}

\author[0000-0002-1046-2685]{Diane Cormier}
\affil{Laboratoire AIM-Paris-Saclay, CEA/DSM-CNRS-Universit\'{e} Paris
 Diderot, Irfu/Service d'Astrophysique, CEA Saclay, Orme des Merisiers, F-91191
 Gif sur Yvette, France}

\author[0000-0001-5414-5131]{Mark E. Dickinson}
\affil{National Optical Astronomy Observatory, 950 North Cherry Avenue,
Tucson, AZ 85719, USA}

\author[0000-0003-0007-2197]{Yu Gao}
\affil{Purple Mountain Observatory/Key Laboratory for Radio Astronomy, Chinese Academy of Sciences, 6 YuanHua Road, Nanjing 210034, People's Republic of China}

\author[0000-0002-8412-7951]{Shuowen Jin}
\affil{Laboratoire AIM-Paris-Saclay, CEA/DSM-CNRS-Universit\'{e} Paris
 Diderot, Irfu/Service d'Astrophysique, CEA Saclay, Orme des Merisiers, F-91191
 Gif sur Yvette, France}
\affil{School of Astronomy and Space Science, Nanjing University,
 Nanjing 210093, China}

\author[0000-0001-5414-5131]{St\'{e}phanie Juneau}
\affil{National Optical Astronomy Observatory, 950 North Cherry Avenue,
Tucson, AZ 85719, USA}

\author[0000-0001-9187-3605]{Jeyhan Kartaltepe}
\affil{School of Physics and Astronomy, Rochester Institute of
Technology, 84 Lomb Memorial Drive, Rochester NY 14623, USA}

\author[0000-0002-9888-0784]{Min-Young Lee}
\affil{Max-Planck-Institut für Radioastronomie, Auf dem Hügel 69, D-53121 Bonn, Germany}

\author[0000-0003-3229-2899]{Suzanne C. Madden}
\affil{Laboratoire AIM-Paris-Saclay, CEA/DSM-CNRS-Universit\'{e} Paris
 Diderot, Irfu/Service d'Astrophysique, CEA Saclay, Orme des Merisiers, F-91191
 Gif sur Yvette, France}

\author[0000-0001-9369-1805]{Annagrazia Puglisi}
\affil{Laboratoire AIM-Paris-Saclay, CEA/DSM-CNRS-Universit\'{e} Paris
 Diderot, Irfu/Service d'Astrophysique, CEA Saclay, Orme des Merisiers, F-91191
 Gif sur Yvette, France}

\author[0000-0002-1233-9998]{David Sanders}
\affil{University of Hawaii, Institute for Astronomy, 2680 Woodlawn Drive, Honolulu, HI, 96822, USA}

\author[0000-0002-0000-6977]{John Silverman}
\affil{Kavli Institute for the Physics and Mathematics of the Universe,
Todai Institutes for Advanced Study, the University of Tokyo,
Kashiwa, Japan 277-8583 (Kavli IPMU, WPI)}

\begin{abstract}

We present the first results of an ALMA survey of the lower fine
structure line of atomic carbon
\cione\ in far infrared-selected
galaxies on the main sequence at $z\sim1.2$ in the COSMOS field. 
We compare our sample with a comprehensive compilation of data available in
literature for local and high-redshift starbursting systems and quasars. We show that
the \ci($^3P_1$$\rightarrow$$^3P_0$) luminosity correlates on global scales with the infrared luminosity
\lir\ similarly to low-$J$ \co\
transitions. We report a systematic variation of
\lprimecifull/\lir\ as a function of the galaxy type, with the ratio
being larger for main-sequence galaxies than for starbursts and
sub-millimeter galaxies at fixed \lir.
The \lprimecifull/\lprimecotwo\ and \mci/\mdust\
mass ratios are similar for main-sequence galaxies and for local and
high-redshift starbursts within a 0.2~dex intrinsic scatter, suggesting that \ci\ is a good tracer of molecular gas mass as \co\ and
dust. We derive a fraction of $f_{\rm{[C\,\scriptscriptstyle{I}\scriptstyle{]}}} = M_{\rm{[C\,\scriptscriptstyle{I}\scriptstyle{]}}} / M_{\rm{C}}\sim3-13$\% of the total carbon
mass in the atomic neutral phase. Moreover, we estimate the \textit{neutral atomic} carbon
abundance, the fundamental ingredient to calibrate \ci\ as a gas tracer,
by comparing \lprimecifull\ and available gas masses
from CO lines and dust emission. We find lower \ci\ abundances in main-sequence
galaxies than in starbursting systems and sub-millimeter galaxies, as
a consequence of the canonical $\alpha_{\rm CO}$ and gas-to-dust conversion
factors. This argues against the application to different galaxy
populations of a universal standard \ci\ abundance derived
from highly biased samples. 

\end{abstract}


\section{Introduction}
\label{sec:introduction}
The growth, structure, dynamics, and eventually fate of star-forming
galaxies is largely regulated by the amount of
gas available to form stars, therefore a crucial parameter to estimate. While low-$J$ transitions of
carbon monoxide $^{12}$CO molecules and optically thin
dust emission are generally trustable tracers for the molecular
gas content of galaxies, they depend on the excitation conditions, metallicity, radiation field,
redshift, geometry, or a good photometric coverage of the far-infrared
emission, and their applicability becomes progressively more observing
time consuming with increasing redshift \citep[e.g.,][]{magdis_2012b, magdis_2017, carilli_2013,
  bolatto_2013, scoville_2014}. Alternative and independent proxies
are, therefore, highly valuable. These include cold
molecular and photodissociation region (PDR) gas tracers, responsible for a
large fraction of cooling (e.g., \ciplus$\,\lambda
158$~$\mu$m, \citealt{zanella_2018}; polycyclic
aromatic hydrocarbon features, PAHs, \citealt{cortzen_2018}).\\

The sub-millimetric atomic carbon transitions
\ci\ ($^3P_1$$\rightarrow$$^3P_0$) ($\nu_{\rm
  rest} =492.161$~GHz) and
\ci\ ($^3P_2$$\rightarrow$$^3P_1$) ($\nu_{\rm rest} =809.344$~GHz) have been put forward as potential
tracers of the bulk of the molecular gas in galaxies. These lines have
an excitation temperature of $T_{\rm ex}=23.6$~K and $62.5$~K,
respectively, and a critical density for collisions with hydrogen atoms
of $n_{\rm crit}\approx10^{3}$~cm$^{-3}$. They can, thus, probe a
wide range of ISM conditions, and they are normally found associated
with PDRs. Early simple plane-parallel modeling of these regions,
predicting \ci\ just in narrow gas slabs between CO and \ciplus,
generated an initial skepticism on the real
usefulness of \ci\ transitions as molecular gas tracers, recently overcome by a growing body of
theoretical and observational work. Modern PDR models including
non-equilibrium chemistry \citep{storzer_1997}, turbulent mixing
\citep{xie_1995, glover_2015}, clumpy geometries \citep{stutzki_1998}, and
the effect of cosmic rays \citep{papadopoulos_2004, bisbas_2015, bisbas_2017} can better explain the
detection of \ci\ fully concomitant with \coone\ and $^{13}$CO over a
wide range of conditions, with a surprisingly constant ratio
$N($\ci$)/N\mathrm{(CO)}\sim0.1-0.2$ and tightly correlated
intensities \citep[e.g.,][]{keene_1996, ojha_2001,
  ikeda_2002}.\\

The use of \ci\ as a tracer of
the molecular gas mass could be even advantageous over the
traditional $^{12}$CO molecule in many respects: (1) the \ci\
lines are as optically thin as $^{13}$CO
($\tau_{\rm{[C\,\scriptscriptstyle{I}\scriptstyle{]}}}\sim0.1-1$,
\citealt{ojha_2001, ikeda_2002} in giant molecular clouds in the Milky
Way and its center), allowing us to probe higher column
densities of cold molecular gas than $^{12}$CO; (2) they do not suffer from the ``excitation bias''
affecting the high-$J$ ($J_{\rm up}\geq 4$) $^{12}$CO transitions, the latter being unable
to capture cool (kinetic temperature $T_{\rm kin}<$ 50~K) and
sub-thermally excited gas (at densities of $n < 10^4$~cm$^{-3}$, \citealt{papadopoulos_2004}),
especially in presence of strong UV radiation fields; (3) models show
that the \cione\
luminosity correlates with the total $M_{\rm gas}$ better than both
ionized carbon \ciplus\ and \coone, regardless of
the local radiation intensity field and spanning at least 4 orders of
magnitude in density ($n = 10 - 10^4$~cm$^{-3}$, Madden et al. in
prep.); (4) for the high
cosmic rays rates expected in high-redshift galaxies, CO is destroyed,
while \ci\ and \ciplus\ become more
abundant \citep{papadopoulos_2004, bisbas_2015, bisbas_2017}; (5) the
simple three-level structure of \ci\ allows
 for breaking the temperature-density degeneracy hampering other
 line tracers, and the excitation conditions of the molecular gas (i.e., its
 excitation temperature $T_{\rm ex}$) can be straightforwardly
 derived from the ratio \lprimecitwo/\lprimeci\
 \citep{weiss_2003}.\\

The \cione\ and \citwo\ transitions have been detected in
molecular clouds of the Galactic disk, the Galactic center, and other
nearby galaxies (\citealt{stutzki_1997, gerin_2000,
  ojha_2001, ikeda_2002} for some
early measurements, \citealt{kamenetzky_2016, israel_2015, lu_2017, jiao_2017} among the
others for more recent compilations). At high redshift, the 
detections reported so far include strongly starbursting
sub-millimeter galaxies (SMGs), radio-galaxies, and quasar hosts (QSOs), often
gravitationally lensed and magnified up to factor of
$30\times$
\citep[e.g.,][and references therein]{walter_2011, alaghband-zadeh_2013,
  gullberg_2016, bothwell_2017, popping_2017}. First results seem to
indicate that in these galaxies the excitation
properties of the interstellar medium and the \ci\ abundances are
similar or more extreme than in (Ultra-)Luminous InfraRed Galaxies
((U)LIRGs) and starbursts in the local Universe. Nevertheless, despite being a
valuable reference sample,
this handful of galaxies ($\sim35$) selected in diverse ways is not
representative of the average
 main-sequence population \citep[e.g.,][]{noeske_2007,
  daddi_2007, elbaz_2007, magdis_2010b}, thus strongly biasing the
general conclusions we can draw about the bulk of high-redshift
galaxies. In order to remedy this situation, we designed and carried out a survey with the
Atacama Large Millimeter Array (ALMA), targeting \cione\
in upper main-sequence
galaxies in the COSMOS field at $z\sim1.2$. Here we present 
the first results on a sample of objects with secure
\cione, dust continuum, and/or \cotwo\ determinations, starting
exploring the potential use of \cione\ as an effective tracer of the
molecular gas on global scales in normal main-sequence galaxies.\\ 

This work is structured as follows. In Section \ref{sec:data} we
describe how we selected the sample of main-sequence galaxies and the
available data; in particular, in Section
\ref{sec:observations} we present the new ALMA data targeting \cione,
along with results from independent programs targeting other \co\
lines for the same objects; in Section \ref{sec:literature} we compile data from the
literature to build a comparison sample for our sources; Section
\ref{sec:results} contains the analysis and the main results of this
work, followed by a discussion in Section \ref{sec:discussion} and the
conclusions in Section \ref{sec:conclusions}.  
Unless stated otherwise, we assume a $\Lambda$CDM cosmology with
$\Omega_{\rm m} = 0.3$, $\Omega_{\rm \Lambda} = 0.7$, and $H_0 = 70$
km s$^{-1}$ Mpc$^{-1}$ and a Chabrier initial mass function
\citep[IMF,][]{chabrier_2003}. All magnitudes are expressed in the AB system.  
All the literature data have been homogenized with our conventions.

\section{Sample and observations}
\label{sec:data}
\begin{figure}
  \centering
  \includegraphics[width=\columnwidth]{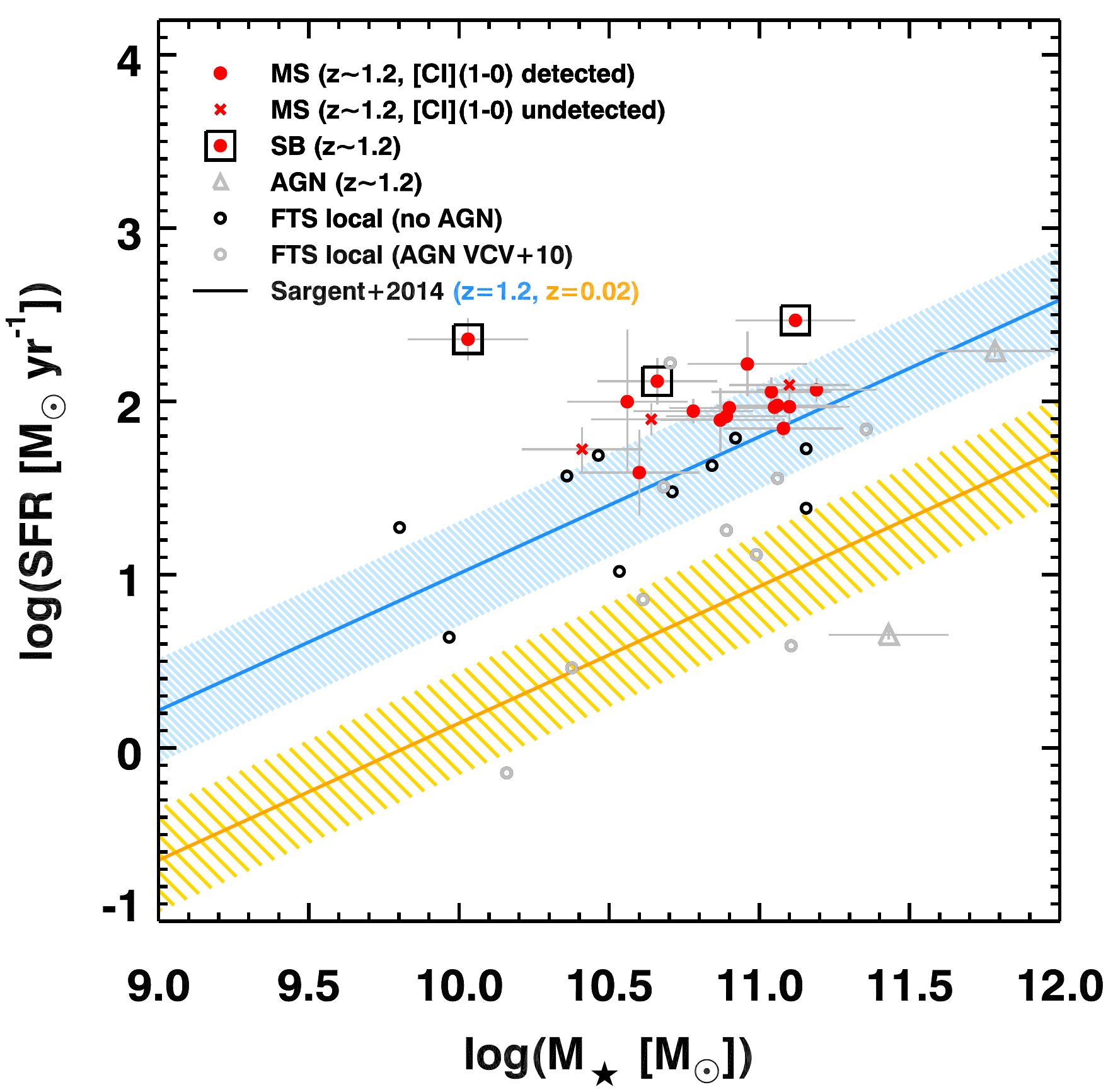}
  \caption{\textbf{Stellar mass - SFR plane.} Red filled circles and crosses
    mark our sample of MS galaxies at $z\sim1.2$ with and without a $>3\sigma$ \cione\
    detection, respectively. Black squares indicate starbursts. Gray triangles mark the
    position of the two AGN dominated galaxy in our sample. Open black
    and gray circles represent
    the local comparison sample of \cione-detected LIRGs and active galaxies, respectively
    \citep{liu_2015}. The blue (orange) solid line and hatched area
    mark the position of the main sequence at $z=1.2$ ($z=0.02$) as parametrized
    by \cite{sargent_2014} and its $1\sigma$ scatter of
    $0.3$~dex.}
  \label{fig:ms}
\end{figure}

\subsection{Sample selection}
\label{sec:sample}
We selected targets in the COSMOS field \citep{scoville_2007} with (1) an available stellar mass
estimate \citep{muzzin_2013, laigle_2016}; (2) a spectroscopic 
confirmation with optical or near-infrared spectrographs from the
COSMOS master catalog (Salvato et al., in prep.); and (3) a \textit{Herschel}/PACS
100 and/or 160~$\mu$m $>3\sigma$ detection in the publicly available
PEP catalog \citep{lutz_2011}. The latter requirement resulted in the selection
of massive galaxies mainly on the upper main sequence and with mean
dust temperatures of $T_{\rm dust} \gtrsim 30$~K. We further chose sources at $z_{\rm opt/NIR}=1.05-1.63$ to
maximize the overlap with parallel and independent ALMA programs targeting \cotwo\ and
\cofive\ (Section \ref{sec:observations}, Daddi et al. in prep.). These criteria drove to an initial
pool of $204$ sources. We then grouped the sources to maximize the
number of targets observable in $3$ frequency configurations of ALMA Band
6. This resulted in the final selection of $50$ sources in the redshift
ranges $z_{\rm opt/NIR}=1.09-1.18$ and $1.23-1.32$, randomly sampling
the whole interval of total infrared luminosities of the
original parent sample. The first redshift
interval allowed us to simultaneously cover \cione\ and the \cofour\ transition
($\nu_{\rm rest} =461.041$~GHz, Section \ref{sec:observations}).\\ 

Here we present the results for $21$ sources with both (1) an
unambiguous spectroscopic confirmation from a sub-millimeter transition,
\textit{and} (2) an estimate of the \cotwo\ flux and/or of the dust
continuum emission, so to ensure at least one molecular gas mass
determination with a standard method (Section \ref{sec:h2_masses}). 
The former criterion allows us to confidently measure even weak
\cione\ fluxes or put stringent upper limits at the expected line
location. In fact, a non detection could be due to either intrinsic weak \cione\
emission or to the absence of frequency coverage owing to inaccurate
redshifts. Significant offsets between optical/near-infrared and
sub-millimeter redshift estimates are not unusual, especially
considering the heterogeneous original data catalogs in the literature and the
different approaches to assess the redshift quality. Some of the
sources we selected did have initial low-quality flags in the COSMOS compilation.
The requirement of alternative gas tracers
excludes $2/29$ extra-sources with a single \cione\ line detection and
no dust continuum or any \co\ emission covered. The rest of
the sample not analyzed here is characterized by: (1) good coverage of the far-infrared
SED, but low quality optical/near-infrared
spectra and no sub-millimeter lines detection ($7/29$ sources) or ascertained wrong
redshifts from \co\ lines that became available after our ALMA
observations ($7/29$ objects). In the latter case, the \cione\ transition fell outside the
covered frequency range or could not be identified unambiguously; 
(2) no detections in $3$ or more followed-up sub-millimeter lines
($4/29$ galaxies), despite good quality flags associated with $z_{\rm
  opt/NIR}$ and good coverage of the far-infrared SED: this might be
due to a wrong association with an
optical/near-infrared counterparts or redshift quality assessment; 
(3) good quality $z_{\rm  opt/NIR}$, but serious blending and source
misidentification in the far-infrared and sub-millimeter bands, which
we could verify only a posteriori with
new catalogs becoming available ($3/29$ sources); (4) a combination of
low quality $z_{\rm  opt/NIR}$ and bad coverage or even non-detection in the far-infrared
SED based on the new catalogs superseding the previous compilations
($6/29$ galaxies). For these galaxies, the absence or a spectroscopic confirmation from a
sub-millimeter line and/or an alternative gas
tracer securely detected, either dust or \cotwo, does not allow a
proper assessment of \cione\ as a proxy for the molecular gas content
in our sample.\\

Respectively $100$\% and $95$\% of the $21$ sources presented here are detected
with a cumulative infrared signal-to-noise ratio $\mathrm{SNR}>3$ and
$>5$ in the ``super-deblended'' 
catalog by \citet[][see below]{jin_2018}. The latter became available after our
ALMA observations and superseded the PEP catalog in our analysis.
Moreover, $16/21$ galaxies lie on the main sequence at their redshift,
$3/21$ are classified as starbursts ($>3.5\times$ above the main
sequence), and $2/21$ suffer from a significant contribution from AGN
emission (Figure \ref{fig:ms}).  The main-sequence galaxies are on
average $\sim1.8\times$ above the 
parametrization by \citealt{sargent_2014}, as expected from our
selection.\\

A discussion of the whole sample and the detailed combined
modeling of all the \co\ and \ci\ transitions is postponed to a
dedicated forthcoming paper (Valentino et al. in prep.).

\subsection{Ancillary data}
\label{sec:ancillary}
Our sample benefits from the excellent photometric and spectroscopic
coverage available in COSMOS. In particular, we adopted the stellar masses listed in
\cite{laigle_2016}, consistent with the values derived by
\cite{muzzin_2013}, both modeling the UV to near-infrared SEDs with standard recipes.
Moreover, we modeled the ``super-deblended''
infrared photometry \cite{jin_2018} as in \cite{magdis_2012}
to derive the galaxy integrated far-infrared properties. The
``super-deblending'' of the highly confused far-infrared bands is
based on an active choice of the radio and
24~$\mu$m priors based on the galaxies spectral properties, reducing the
blending degeneracies and resulting in well behaved flux density
uncertainties (see \citealt{liu_2018} for a detailed description of
the method).
Whenever available, we fit the
emission from \textit{Spitzer}/MIPS 24~$\mu$m \citep{sanders_2007},
\textit{Herschel}/PACS \citep{lutz_2011} and SPIRE bands
\citep{oliver_2012}, JCMT/SCUBA2 \citep{geach_2017}, ASTE/AzTEC
\citep{aretxaga_2011}, IRAM/MAMBO \citep{bertoldi_2007}, and ALMA continuum
emission at $\sim1.1-1.3$~mm (Section \ref{sec:observations}) with an expansion of the
\citet[DL07]{draine_2007}
model library (Figure \ref{fig:data} in Appendix). We further included a dusty torus component
surrounding active galactic nuclei (AGN) following \cite{mullaney_2011} and subtracted
this contribution from the total \lir\ we derived. Therefore, \lir\
always refers only to the component due to star formation for this
sample. We then converted \lir\ into SFR as $\mathrm{SFR} =
L_{\mathrm{IR}}\,[L_\odot] / (9.86 \times 10^9)$~\myr\
(\citealt{kennicutt_1998}, converted to a Chabrier IMF).
The emission from
the dusty torus is relevant ($\sim40$\% and $>95$~\% of \lir) in $2$ sources,
flagged everywhere as ``\textsc{AGN}'' hereafter. Moreover, their
stellar masses are likely overestimated, due to significant AGN emission in
the near-infrared bands. Therefore, for the
purpose of this work, we will not include the AGN in any further step
of the analysis.
While fitting the SEDs, we included the upper limits in every band, modeling the nominal values
weighted by their large uncertainties. We then bootstrapped the values within the
observed errors to estimate the statistical uncertainties on the
derived quantities. In Table \ref{tab:fir} and \ref{tab:mgas}, we report the
$8-1000$~$\mu$m total \lir\ for all our galaxies, the contribution from dusty tori, and the total dust mass \mdust. 
We note that \lir\ is
well constrained for the vast majority of our sample, while \mdust\ critically relies on the availability of a
measurement in the Rayleigh-Jeans tail of the dust
emission.

\subsection{ALMA [C\,{\footnotesize I}](1-0) and CO observations}
\label{sec:observations}
We collected Atacama Large Millimeter Array (ALMA) Band 6 observations
during Cycle 4 (Project ID: 2016.1.01040.S, PI:
F. Valentino). Galaxies were grouped in three scheduling blocks
targeting \cione\ ($\nu_{\rm rest} =492.161$~GHz)
and \cofour\ ($\nu_{\rm rest} =461.041$~GHz) at $z\sim1.15$, and \cione\ only at $z\sim1.28$ within
contiguous spectral windows (SPWs) of 1.875~GHz and a requested spectral
resolution of $7.8$~MHz ($\sim10$~\kms). Two out of three of the blocks
were fully observed, while the third was incomplete, resulting in a
higher rms. Data were collected in
configuration C40-1, corresponding to a synthesized beam of
$2.0\times1.7$''. Galaxies are generally not (or marginally) resolved,
ensuring minimal flux losses. We
reduced the raw data with the standard ALMA pipeline with CASA
\citep{mcmullin_2007}. We then converted the calibrated data cubes to
\textit{uvfits} and analyzed them with GILDAS
\citep{guilloteau_2000}. We extracted 1D spectra using PSFs and circular Gaussians
models and fitting visibilities in the \textit{uv} space with the iterative
process described in \cite{daddi_2015}. The \cione\ spectra are shown in
Figure \ref{fig:data} in Appendix. We looked for emission
lines scanning the signal-to-noise ratio (SNR) spectrum. We measured fluxes as the weighted
average flux density within the channels maximizing the SNR, times the
velocity width covered by these channels. We
further fit single or double Gaussians to the line profile to estimate
total fluxes, generally $\sim10$\% larger than the fluxes measured
over the number of channels maximizing the SNR. We finally adopted
the first approach, applying a correction of $10$\% to the line fluxes. When multiple lines
were available, we measured fluxes and upper limits on the same velocity width of the
brightest line. These results agree with measurements leaving
each line center and width free to vary. We measured integrated
\cione\ fluxes at $>3\sigma$ in $18/21$ sources down to an
average rms/beam of $\sim0.15$~\jykms\ for a line width of $400$~\kms\ and a
final velocity resolution of $\sim20-40$~\kms. Fifteen out of $21$
sources have \ci\ detections significant at
$>4\sigma$. All the remaining sources have either one or multiple CO lines
detected at $>4\sigma$ at the same redshift (see next paragraph), allowing us to explore
the $\mathrm{SNR}<4$ regime or to put secure upper limits on \ci.
We simultaneously measured
the continuum emission at observed $\sim1.3$~mm over $7.5$~GHz assuming an
intrinsic slope of $\nu^{3.5}$ ($\beta=1.5$). We detected significant continuum emission at
$3\sigma$ in $14/21$ sources down to an rms of $\sim0.07$~mJy on the
full frequency range.\\

From the same observing campaign, we similarly measured \cofour\
fluxes at $>4\sigma$ significance in all $14$ galaxies with frequency
coverage of this line. Moreover, 15
and 11 galaxies of our sample have been observed by ALMA Bands 6 and 3
independent observations
targeting \cofive\ and \cotwo, respectively (Project IDs: 2015.1.00260.S,
2016.1.00171.S, PI: Daddi; Daddi et al. in prep.). Data were
collected at similar spatial resolutions, reduced and analyzed as
we described above. For the purpose of the present work, we used
high-SN CO emissions to (1) fix the center of the
circular Gaussian or PSF to extract the spectrum, the central line
frequency, and the width of \cione; (2) to
measure gas masses from \cotwo; and (3) to measure mm continuum
emission. We do not find evidence of systematically broader or narrower \ci\ lines 
than CO transitions, when the velocity width is let free to vary.
We significantly detected \cotwo\ and \cofive\ in all 11 and 15 targeted galaxies, and
continuum emission at observed $\sim3-1.1$~mm for $1/11$ and $8/15$
galaxies covered by Band 3 and 6 observations, respectively.
This brings the overall number of sources with $\sim1"$ spatial resolution mm continuum detection
to $17/21$.\\

We report the observed \cione\ and \cotwo\ fluxes and $L'$ luminosities in Table
\ref{tab:fir}. A full compilation including the \cofour\ and \cofive\
fluxes will be discussed in future work (Valentino et al. in prep.).

\section{Literature data}
\label{sec:literature}
We compared our results with observations available in literature. We
collected and homogenized line luminosities and galaxy properties to
properly match our conventions. 

\subsection{Local galaxies}
\label{sec:lowzsample}
This sample is drawn from the public compilation of all
\textit{Herschel}/Fourier Transform Spectrometer (FTS) observations in the
\textit{Herschel} Science Archive of local galaxies by
\citet[L15 hereafter]{liu_2015}. These sources are part of the \textit{IRAS}
Revised Bright Galaxy Sample \citep{sanders_2003} and covered at
$70-160$~$\mu$m by \textit{Herschel}/PACS. FTS simultaneously spanned
the $446-1543$~GHz frequency interval, covering all CO lines with
$J_{\rm up}=4 - 13$, the \cione\ and \citwo\ lines, and several
other transitions. L15 reduced the FTS raw data with the SPIRE v.12 calibration
products and the \textit{Herschel} Interactive Processing Environment
pipelines \citep[HIPE
v12.1.0,][]{ott_2010}. They extracted all the lines
simultaneously with customized optimized HIPE spectral line fitting scripts
on the unapodized spectra with varied-width Sinc-convolved Gaussian
(SCG) functions, and they estimated the line flux errors from the rms of the
spectra near each line (see L15 for further details).
In total we retrieved $32$ galaxies (out of $146$ in the compilation by L15) with a
\cione\ $>3\sigma$ detection up to $z\sim0.03$. 
We corrected the far-infrared luminosities integrated over the
$40-400$~$\mu$m interval reported in
L15 to match the $8-1000$~$\mu$m total
\lir\ we adopt here, by multiplying by a factor $1.2\times$. We
obtained this value by comparing the original $L_{\rm
  FIR}(40-400\,\mu\mathrm{m})$ with $L_{\rm
  IR}(8-1000\,\mu\mathrm{m})$ from SED modeling as described in Section
\ref{sec:ancillary} for the subset of galaxies from the Great
Observatories All-Sky LIRGs Survey \citep[GOALS,][]{armus_2009}
included in L15. Emission line fluxes and infrared
  luminosities have been beam-matched as described in
L15 and reported to the total, galaxy-integrated values back-applying
the beam correction based on PACS photometry
($I_{_{\rm{[C\,\scriptscriptstyle{I}\scriptstyle{]}}}}(\mathrm{L15,
  total}) =
I_{_{\rm{[C\,\scriptscriptstyle{I}\scriptstyle{]}}}}(\mathrm{L15,
  beam})\times F_{\rm PACS, total}
/ F_{\rm PACS, beam} $).\\

We further cross-matched the sample in L15
with the alternative compilation of \textit{Herschel}/FTS observations
and low-$J$ CO transitions from ground based facilities by
\citet[K16]{kamenetzky_2016}. First, we checked that the FTS beam measurements
for galaxies with \cione\ detections in both samples were consistent.
Then, we corrected the CO line measurements within the fixed 43.5''
beam in K16 (their Table 3) to the
galaxy-integrated values. We multiplied the fluxes in K16 by $\eta_{\rm \,beam} = I_{\rm{[C\,\scriptscriptstyle{I}\scriptstyle{]}}}(\mathrm{L15,
  total})/I_{\rm{[C\,\scriptscriptstyle{I}\scriptstyle{]}}}(\mathrm{K16,
  43.5" beam})$, where $I_{\rm{[C\,\scriptscriptstyle{I}\scriptstyle{]}}}(\mathrm{L15,
  total})$ are the galaxy-integrated \cione\ fluxes in L15 and $I_{\rm{[C\,\scriptscriptstyle{I}\scriptstyle{]}}}(\mathrm{K16,
  43.5" beam})$ the \cione\ fluxes within a fixed 43.5'' beam by K16. The median correction factor is $2\times$ and
$1.25\times$ for sources closer and farther than $20$~Mpc,
respectively. When multiple estimates of the same low-$J$ \co\
transitions were available, we assumed a SNR-weighted average as
representative of the line flux.
Note that the line ratios do not suffer from extra uncertainty due to the
beam correction than what reported in the K16 compilation. Moreover,
for the closest sources the final values might be representative mainly of the nuclear regions.
Out of $32$ sources with a \cione\ detection, $29$ and $26$ have a \coone\
and \cotwo\ $>3\sigma$ detection, respectively.\\

We further estimated the stellar masses of $20/32$ galaxies of the
FTS sample with available $K_{\rm s}$-band imaging from 2MASS
\citep{skrutskie_2006} averaging the values obtained following
\citet[Eq. 2]{arnouts_2007} and \citet[Eq. B1,
B2]{juneau_2011}. We checked these results against the full UV to
near-infrared SED fitting for a subset of $32$
objects in common between the whole compilation of L15 and the sample
studied by \cite{u_2012}, finding
consistent results. 
Figure \ref{fig:ms} shows the location of the galaxies with a $>3\sigma$
\cione\ detection in the \mstar\ -- SFR(\lir) plane, typically lying above the
main sequence at their redshift. 
Since AGN can contaminate both \lir\ and the
$K_{\rm s}$-band derived \mstar, we flagged $12/32$ known active galaxies
listed in the catalog by \citet[VCV10]{veron-cetty_2010}. Among the
non-active galaxies, $70$\% ($14/20$) are LIRGs ($L_{\rm IR}>10^{11}$~\lsun).
For the local sample, we do not attempt to disentangle the
contribution of the dusty torus to the total \lir. 
In the rest of the paper, we will refer to the sample of local
galaxies as ``FTS Local'' or ``Local LIRGs'', when they do not host a bright AGN.
\begin{figure*}
  \centering
  \includegraphics[width=\textwidth]{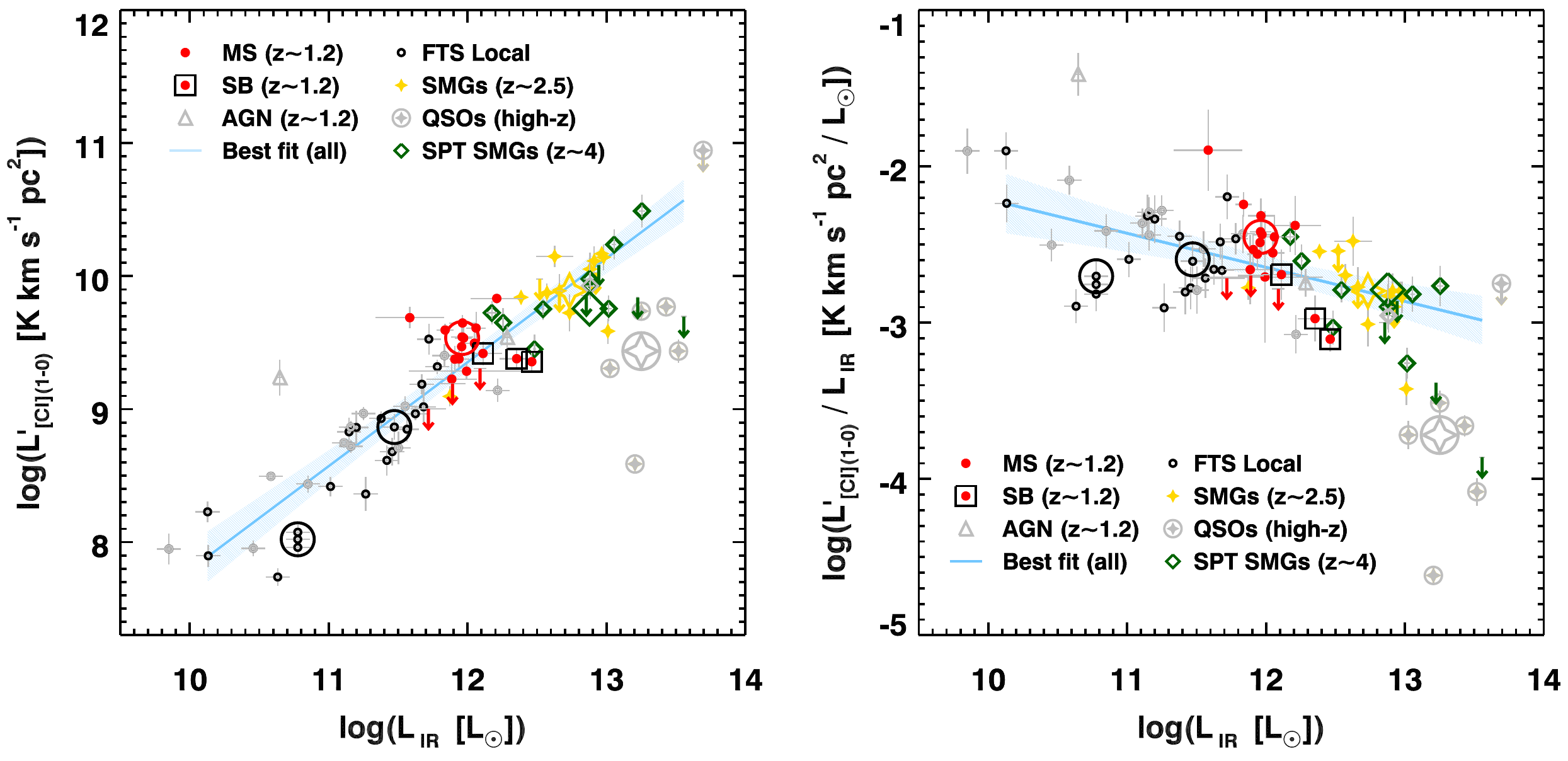}
  \caption{\textbf{Total \lir\ - \lprimeci\ relation.} In both panels,
  red filled circles represent our $>3\sigma$ detected MS galaxies at
  $z\sim1.2$; open black squares: SBs; gray triangles:
  AGN; open black circles: local FTS $>3\sigma$ detections without AGN
  signatures; open gray circles: local FTS $>3\sigma$ detections
  listed as active in VCV10; yellow
  stars: SMGs at $z\sim2.5$ from W11 and AZ13; gray stars and open circles:
  high-redshift QSOs from W11; green open diamonds: SPT SMGs at
  $z\sim4$ from B17. Arrows mark $3\sigma$ upper limits on
  \lprimeci. The solid light blue line and hatched area represent the
  best linear regression model of all SF-dominated galaxies (i.e.,
  excluding QSOs and AGN) and the $95$\% confidence interval.
  The \lir\ of QSOs is dominated by the emission of the dusty
    torus, which is included in the estimate, at odds with the rest of
  SF-dominated galaxies.
  The larger symbols mark the median values for each sample. We split
  the FTS local sample in two bins with a threshold at \lir$=10^{11}$~\lsun, the
  formal limit for the LIRG classification.
  \textit{Left:}
    \lir\ -- \lprimeci\ relation, a proxy for the SFR and gas mass
    relation. 
   \textit{Right:} \lir\ -- \lprimeci/\lir\ relation, where
   \lprimeci/\lir\ is a proxy for the gas depletion timescale $\tau_{\rm
     dep}\propto 1/\rm{SFE}$, dropping with increasing \lir\ and
   shorter for starbursting galaxies than for sources on the MS.}
  \label{fig:lir_lprimeci}
\end{figure*}

\subsection{High-redshift sub-mm galaxies and quasars}
\label{sec:highzsample}
We assembled a sample of sub-millimeter galaxies (SMGs) and quasars (QSOs)
with \cione\ coverage from \citet[W11]{walter_2011}, \citet[AZ13]{alaghband-zadeh_2013}, and
\citet[B17]{bothwell_2017}. We refer the reader to these papers for
fully detailed references, sample selection, and
observations. W11 and AZ13 targeted or collected information on
typical SMGs at $z\sim2.5-4$ detected at $\sim850$~$\mu$m, with a tail
of well-studied QSOs extending up to $z\sim6.5$. A large fraction of
these sources are gravitationally magnified up to $\sim30\times$ and are detected in
\cothree\ and \cofour. Similarly magnified are
$1.4$~mm detected sub-millimeter galaxies in B17, identified in a
blank-field survey with the South Pole Telescope (SPT,
\citealt{vieira_2010, weiss_2013}), with spectroscopic information on
high- \citep[B17]{weiss_2013} and low-$J$ transitions \citep[\cotwo,][]{aravena_2016}.
 The final sample consist of $33$ galaxies,
$25/33$ with \cione\ detections at $>3\sigma$. We re-derived their total
\lir\ and dust masses \mdust,
modelling their far-infrared SED applying the same method described in
Section \ref{sec:ancillary}. The SPT-SMGs galaxies are all detected in
SPIRE $250$, $350$, and $500$~$\mu$m, LABOCA $870$~$\mu$m, SPT
1.4 and 2.0~mm, and ALMA 3~mm bands \citep{weiss_2013}. The SEDs
of SMGs and QSOs from W11 and AZ13 are sampled less homogeneously, but
ensuring a good coverage of both the peak and the Rayleigh-Jeans of
the dust emission in the vast majority of cases. Applying our recipes,
we estimate total
$L_{\rm{IR}}(8-1000\,\mu m)$
$\sim1.5\times$ ($\sim3\times$) larger than the original values
derived with modified
black body curves for SMGs (QSOs), including a correction for the different integration limits. We also estimate
\mdust\ for the SPT-SMGs systematically larger than previously reported
\citep{aravena_2016}. The systematic differences in \lir\ and
  \mdust\ fully depend on the adopted models (modified black body law,
  DL07) and their parameters (effective dust emissivity index $\beta$, dust mass absorption
  coefficient $\kappa$, peak temperature,
  \citealt{magdis_2012}). The discrepancy in \lir\ is larger for QSO hosts
  owing to the dusty torus emission mid-IR bands,
  where the difference between modified black body curves and
  DL07 models is more significant.
All the quantities presented here have
been corrected for magnification. Moreover, we correct the \lir\ luminosities of SMGs
for the contribution of AGN similarly to what we did for the MS
sample. We find only $1$ (SPT-)SMG whose SED is dominated by a dusty
torus ($\sim70$\% of the total \lir). For known bright QSOs at high-redshift in W11, we
do not attempt to separate the star formation and AGN contributions to
\lir, being largely dominated by the latter. However, we will not consider these sources in
the analysis any further, but simply show their position in the
various plots for reference. Stellar masses are
not available for this high-redshift sample, apart from $2$ sources listed in
B17. Therefore, we could not place these objects in the \mstar\ -- SFR
plane and canonically define them as SB or MS based on these
observables. However, their observed ISM conditions, gas and SFR
densities, and SFEs generally distinguish SMGs from MS galaxies
\citep[e.g., ][]{daddi_2010, genzel_2010, bothwell_2013,
  casey_2014}. In the following, we will label ``SMGs ($z\sim2.5$)''
the sample from W11 and AZ13, ``QSOs (high-z)'' the sources with clear
AGN signatures from W11, and ``SPT SMGs ($z\sim4$)'' the sample by
B17. Moreover, we will consider SMGs as starbursting systems and not
typical MS galaxies.
\begin{figure}
  \centering
  \includegraphics[width=\columnwidth]{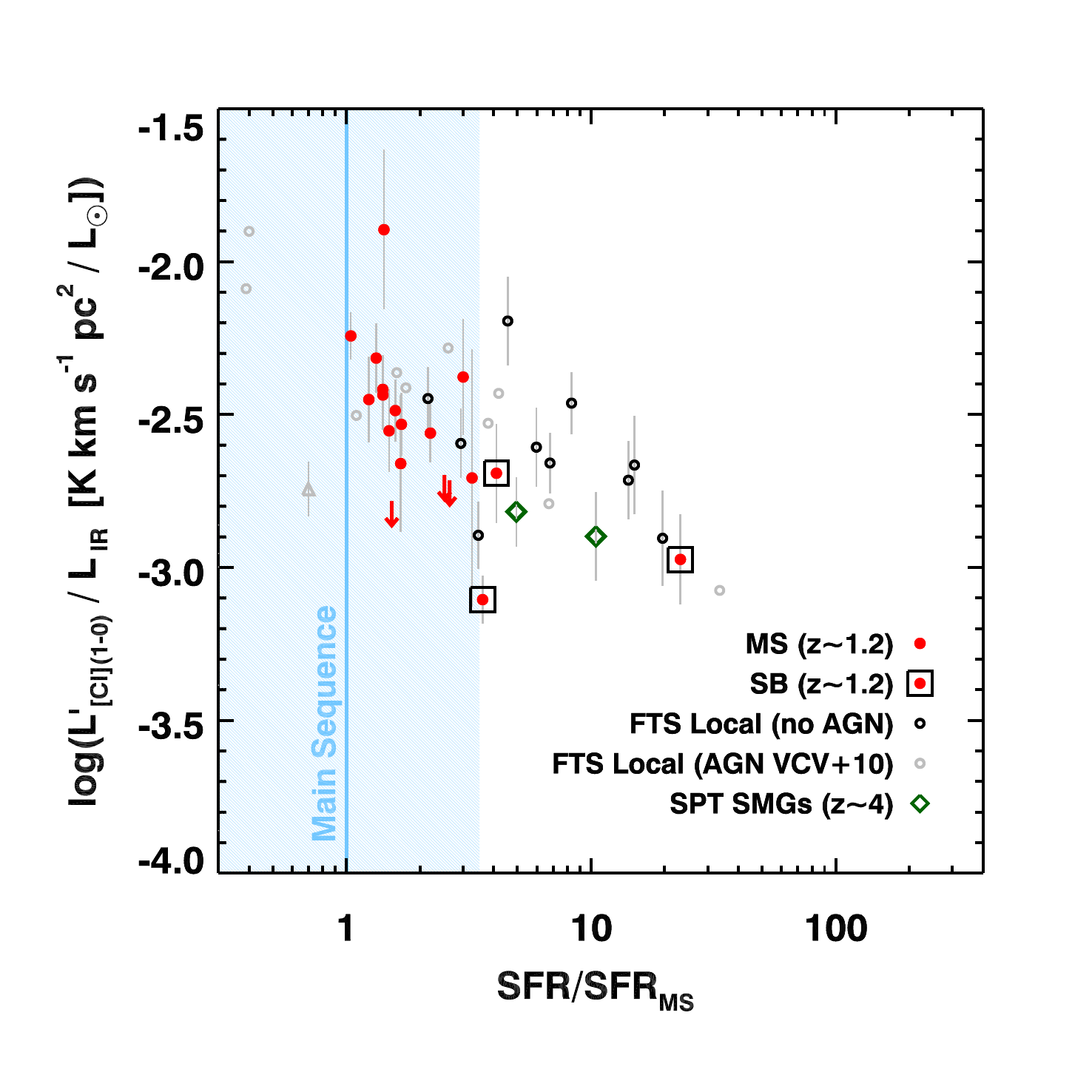}
  \caption{\textbf{Distance from the main sequence - \lprimeci/\lir\ plane.} Red filled circles
    mark our sample of MS galaxies at $z\sim1.2$ with a $>3\sigma$ \cione\
    detection. Black squares indicate SBs. Open black and gray circles represent
    the local comparison sample of non-active and active
    galaxies covered by FTS observations
    \citep{liu_2015}. Green open diamonds indicate the $2$ SPT-SMGs
    at $z\sim4$ with an estimate of \mstar\ from B17. Arrows mark $3\sigma$ upper limits on
  \lprimeci. The MS is parametrized as in \cite{sargent_2014}.}
  \label{fig:dsfr}
\end{figure}

\section{Analysis and results}
\label{sec:results}

\subsection{The \lir\ -- \lprimeci\ relation}
\label{sec:lir_lprimeci}
Following \cite{solomon_2005}, we computed the \cione\ line
luminosities in \kkmspc, representing the integrated source brightness temperature:
\begin{equation}
L'_{\rm line} \, [\mathrm{K\,km\,s^{-1}\,pc^2}]= 3.25 \times 10^{7}\,
S_{\rm line}\, \Delta v \, \nu_{\rm
  obs}^{-2} (1+z)^{-3} D_{\rm L}^2
\end{equation}
where $S_{\rm line}\,\Delta v$ is the measured velocity-integrated line
flux in Jy$\,$km$\,$s$^{-1}$, $\nu_{\rm obs}$ the
observed line frequency in GHz, $z$ the redshift of the
source, and $D_{\rm L}$ is the luminosity distance in Mpc. 
Figure \ref{fig:lir_lprimeci} shows the relation between the total
\lir\ and \lprimeci\ for our sample of MS galaxies and the literature
compilation of starbursting sources at various redshifts. Both a Spearman's
rank and a Pearson's correlation coefficients show that the two
quantities are correlated, considering all $>3\sigma$ detected
sources and excluding QSOs and AGN ($\rho_{\rm Spearman}=0.8990$,
$\rho_{\rm Pearson}=0.9024$). 
We further applied a linear regression analysis on
$\mathrm{log}(L_{\rm IR})$ -- $\mathrm{log}($\lprimeci$)$ both using
a Bayesian (\textsc{linmix\textunderscore err.pro},
\citealt{kelly_2007}) and a $\chi^2$-minimization algorithm (\textsc{mpfit.pro}, \citealt{markwardt_2009})
taking into account the uncertainties on both \lir\ and \lprimeci, and
including the upper limits on \lprimeci\ in the Bayesian fit. Since \lir\ and
\lprimeci\ are proxies for the integrated SFR and \mgas, this relation
is analogous to the Schimdt-Kennicutt relation \citep[][with the X and
Y axes generally inverted]{schmidt_1959,
  kennicutt_1998b}. The two algorithms we applied provided fully
consistent results within the uncertainties,
and the effect of upper limits is negligible. 
We modeled a total of
$57$ \cione-detected galaxies and $10$ upper limits.
The Bayesian best fit model returns a sublinear slope of $0.78\pm0.05$ with an observed
scatter of $\sigma=0.26$~dex. Note that AGN and QSOs are not included in
the fit, nor in the calculation of $\sigma$. Their location in
the diagram is mainly driven by their \lir, boosted by the
contribution of the dusty tori in the mid-IR regime, adding to
moderately larger intrinsic luminosities than high-redshift SMGs at fixed \lprimecifull\ (W11).
Modeling only the
starbursting population (i.e., SB at $z\sim1.2$, local LIRGs, and SMGs)
provides a similar slope of $0.79\pm0.06$.
Interestingly, these values are consistent with that of the
$\mathrm{log}(L_{\rm IR})$ -- $\mathrm{log}(L'_{\rm{CO(1-0)}})$ relation
($0.81\pm 0.03$, \citealt{sargent_2014}), reinforcing the connection
between \ci\ and CO. Moreover, our MS detected galaxies appear to
have larger \lprimeci/\lir\ ratios than SMGs. This is more
evident in the right panel of Figure
\ref{fig:lir_lprimeci}. The mean value of
  $\mathrm{log}(L'_{\mathrm{[C\,\scriptscriptstyle{I}\scriptstyle{]}}^3P_1\,-\,
    ^3P_0}/L_{\mathrm IR} \, [\mathrm{K\,km\,s^{-1}\,pc^{2}\,L_\odot^{-1}}])$ for $16$ MS galaxies is
$(-2.43\pm0.06)$, $\sim2\times$ higher than the
mean for SMGs at $z\sim2.5$ ($(-2.77\pm0.07)$~dex) and
SPT-SMGs at $z\sim4$ ($(-2.80\pm0.07)$~dex), where
the uncertainties represent the error on the mean. We included the
upper limits on \cione\ in the calculation using a survival analysis
\citep[KM estimator,][]{kaplan-meier_1958}. This 
difference is significant at a $\sim4\sigma$ level. The median
values are fully consistent with the mean. The ratio for the
local sample of non-active galaxies is consistent with the estimate for MS galaxies, but
it suffers from a very large dispersion. In Figure \ref{fig:dsfr} we further
show the \lprimeci/\lir\ ratio as a function of the distance from the
main sequence ($\Delta \mathrm{SFR=SFR/SFR_{\rm MS}}$) as parametrized
in \cite{sargent_2014}. We included only sources with a stellar mass
estimate, i.e., all our galaxies, part of the local LIRGs,
and $2$ SPT-SMGs. Excluding galaxies with AGN signatures, the \lprimeci/\lir\  ratio and $\Delta \mathrm{SFR}$
are mildly anti-correlated ($\rho_{\rm Spearman}=-0.6940$, $\rho_{\rm
  Pearson}=-0.5961$), with SMGs and
SBs at $z\sim1.2$ showing systematically lower ratios than MS
galaxies, as in Figure \ref{fig:lir_lprimeci}. 
However, the scarce statistics of lower main-sequence sources and SBs
with available \mstar\ prevents us from deriving more definitive
conclusions. From a physical perspective, since \lprimeci\ traces
the gas mass and \lir\ the SFR, their ratio is a proxy for the
gas depletion timescale $\tau_{\rm dep}$. The observed trends would then
suggests a drop of this quantity (or equivalently an increment of SFE)
with increasing \lir\ and distance from the main sequence,
analogously to the well established correlations observed for
  CO \citep[e.g.,][]{daddi_2010b, magdis_2012, genzel_2015, tacconi_2018}. 

\subsection{The \lprimeci\ -- \lprimecotwo\ relation}
\label{sec:lprimeci_lprimecotwo}
\begin{figure*}
  \centering
  \includegraphics[width=\textwidth]{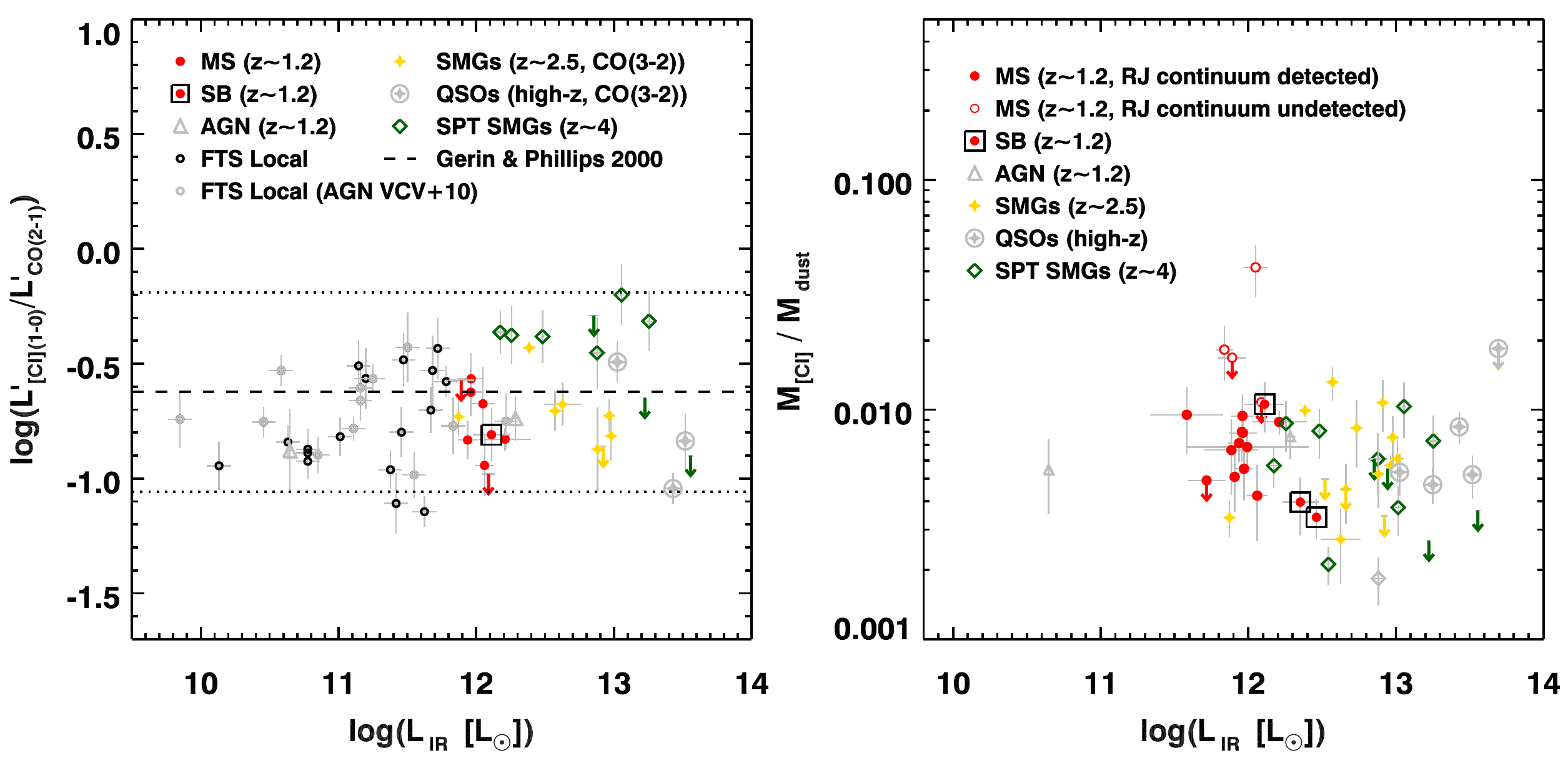}
  \caption{\textit{Left:} \textbf{\lprimeci\ -- \lprimecotwo\ as a function of total
      \lir.} Symbols are coded as in Figure \ref{fig:lir_lprimeci} and
    reported in the legend. The dashed and dotted black lines mark the average
    ratio and its scatter observed in a sample of local spirals, mergers, and
    low-metallicity galaxies from \cite{gerin_2000}. Arrows mark
    $3\sigma$ upper limits on \lprimeci. \textit{Right:}
    \textbf{\mci/\mdust\ as a function of \lir.} Symbols are
    coded as in Figure \ref{fig:lir_lprimeci} and
    reported in the legend. Open red circles mark sources without a
    reliable continuum detection in the Rayleigh-Jeans tail, thus
    having an uncertain dust mass. In both panels, the \lir\
      of QSOs includes the emission of the dusty tori, at odds with the rest of
      SF-dominated galaxies.}
  \label{fig:ci_co_dust}
\end{figure*}
Figure \ref{fig:ci_co_dust} shows the
\lprimeci/\lprimecotwo\ ratio as a function of the total \lir. The
observed ratio is similar in MS and SB galaxies at $z\sim1.2$, local
LIRGs, and high-redshift SMGs, within a fairly large
scatter. We estimate a mean value of
$\rm{log(}$\lprimeci/\lprimecotwo$)=(-0.69\pm0.04)$ with an observed
scatter of $0.23$~dex for $37$ $>3\sigma$-detected galaxies, largely dominated by the intrinsic dispersion
of $0.20$~dex. The inclusion of $6$ upper limits with a survival
analysis provides consistent values (the mean from the
\citet{kaplan-meier_1958} estimator is
$\rm{log(}$\lprimeci/\lprimecotwo$)=(-0.72\pm0.04)$, with a standard
deviation of 0.24~dex).
The \lprimeci/\lprimecotwo\ ratio does not appear to vary with \lir. For the W11
sample we converted \cothree\ to \cotwo\ applying a $r_{\rm
  32}=L'_{\rm CO(3-2)}/L'_{\rm CO(2-1)}=0.62$
\citep{bothwell_2013}. Adopting ratios close to the original W11
paper ($r_{32}=1$ for fully thermalized gas) would result in ratios $\sim60$\% larger,
bringing them closer to the values from B17. For the FTS sample we used the observed
\cotwo\ luminosities when available ($26/32$ galaxies), and converted
\coone\ to \cotwo\ fixing $r_{\rm
  21}=L'_{\rm CO(2-1)}/L'_{\rm CO(1-0)}=0.84$ for $3$ more sources.
These ratios are consistent with the observations in
local spirals, mergers, and low-metallicity galaxies and their large scatter reported in
\cite{gerin_2000}, once corrected for the small excitation bias
between \cotwo\ and \coone\
($\rm{log(}$\lprimeci/\lprimecotwo$)=-0.7\pm0.4$~dex, with extreme
values of $\rm{log}($\lprimeci/\lprimecotwo$)\sim-1.4$ and $0$). 
Therefore, \cione\ and low-$J$ \co\
transitions appear to be correlated on $>$kpc-scales regardless of
galaxy type, total \lir, and redshift, although with substantial
scatter due to object by object variations, given the small
measurement errors on fluxes.

\subsection{Neutral atomic carbon masses}
\label{sec:ci_masses}
Following \cite{weiss_2005}, the mass of atomic carbon is derived straightforwardly from \lprimeci\
as: 
\begin{equation}
M_{\rm{[C\,\scriptscriptstyle{I}\scriptstyle{]}}}  = 5.706\times10^{-4} \,
Q(T_{\rm ex}) \frac{1}{3}\, \mathrm{e}^{23.6/T_{\rm ex}} \, L'_{\mathrm{[C\,\scriptscriptstyle{I}\scriptstyle{]}}^3P_1\,-\, ^3P_0}\,[M_\odot]
\end{equation}
where $Q(T_{\rm ex}) = 1 + 3\mathrm{e}^{-23.6\rm{K}/T_{\rm ex}} +
5\mathrm{e}^{-62.5\rm{K}/T_{\rm ex}}$ is the partition function of \ci\ and $T_{\rm
ex}$ the excitation temperature. We cannot derive $T_{\rm ex}$ from the
\lprimecitwo/\lprimeci\ ratio, since \citwo\ is unavailable for our sample of MS
galaxies at $z\sim1.2$. Therefore, we assume a fixed \ci\
excitation temperature $T_{\rm ex} =
30$~K for all galaxies. W11 reports a $\langle T_{\rm ex}
  \rangle=29.1 \pm 6.3$ K for their overall sample of SMGs and QSOs and we
  derive $\langle T_{\rm ex} \rangle
= 25 \pm 1$~K for part of the FTS local sample with both \cione\ and
\citwo. A typical value of $30$~K has also been previously adopted by AZ13
and B17. Assuming the 
dust temperature that we estimate from the far-infrared SED modeling 
as a first guess for the \ci\ $T_{\rm ex}$, we find similar results
($\langle
T_{\rm dust} \rangle = 31 \pm 1$~K for the MS calibration sample, and $\langle
T_{\rm dust} \rangle = 36 \pm 2$~K for the SBs). Note that the total
neutral carbon mass is insensitive to $T_{\rm ex}$ above
$20$~K \citep{weiss_2005}, so the exact choice of this parameter is
not critical above this threshold, which is unlikely in the samples studied here. We report the total \mci\ masses in Table \ref{tab:mgas}.

\subsection{Gas masses from dust and CO}
\label{sec:h2_masses}
We estimated the total gas masses (including a $1.36\times$
contribution from helium) for our sample of MS galaxies using both
dust masses from SED modeling and \coone\ or \cotwo\ as gas tracers, when
available. In both cases we adopted a
metallicity-dependent conversion
factor as described in \cite{magdis_2012}\footnote{\citet{magdis_2012}
  reported an intercept of 12.8 in their parametrization
  of $\alpha_{\rm CO} (12+\mathrm{log(O/H)})$ (Eq. 8), not matching their Figure 5 (right). The
  correct value adopted here is 12.6. The parametrization of
  $\delta_{\rm GDR}(12+\mathrm{log(O/H)})$ is
  identical to Magdis et al.}. Note that this approach includes the
atomic hydrogen \hi\ in the gas mass estimate, a significant
contributor to the total gas mass only at low redshift.
We derived metallicities
converting stellar masses and SFR with the Fundamental
Metallicity Relation \citep{mannucci_2010}.
The derived metallicities are generally consistent with the solar
value ($12+\rm{log(O/H)_{\odot}} = 8.69$, \citealt{asplund_2009}). 
We estimate an average gas-to-dust conversion factor of
$\langle \delta_{\rm GDR} \rangle \sim 87 $ and $\langle \alpha_{\rm CO}
\rangle \sim 3$~\msun/\kkmspc. We assumed a
\lprimecotwo/\lprimecoone\ ratio of $r_{21} = 0.84$
to convert \cotwo\ into total gas masses when necessary \citep{magdis_2012, bothwell_2013}.
We further derived total gas masses for our SBs at $z\sim1.2$, LIRGs and high-redshift
SMGs fixing the conversion factors to $\alpha_{\rm
  CO}=0.8$~\msun/(\kkmspc), and $\delta_{\rm GDR}=30$. The final error budget includes the uncertainties
on the observed \co\ fluxes and dust mass from SED modeling (Section
\ref{sec:ancillary}). We further include a $0.2$~dex statistical
error on $\alpha_{\rm CO}$ and $\delta_{\rm GDR}$, mimicking
the uncertainty on the metallicity-dependent parametrization in
\cite{magdis_2012}. Possible larger systematic uncertainties affecting
the gas masses are not listed in the error budget (e.g.,
\citealt{kamenetzky_2017} for a study of the local LIRGs).\\

Figure \ref{fig:ci_co_dust} shows that the \mci/\mdust\ ratios of
SMGs appear similar to the values for our MS and SB galaxies at
$z\sim1.2$, albeit with substantial scatter. We estimate a mean ratio
of $\mathrm{log(}$\mci/\mdust$)=(-2.20\pm0.03)$, with an observed
scatter of $\sigma=0.19$~dex dominated by an intrinsic dispersion of
$0.15$~dex for 33 $>3\sigma$-detected galaxies. The inclusion of  $11$ upper limits
with a survival analysis provides a consistent result
($\mathrm{log(}$\mci/\mdust$)=(-2.26\pm0.04)$, $\sigma=0.23$~dex). Note that we excluded active galaxies, QSOs, and galaxies without a detection of
the dust continuum from this calculation. 
Moreover, the SMGs
from B17 at $z\sim4$ appear to have
fainter \cotwo\ emission than the sample from W11 at $z\sim2.5$ at
fixed dust mass, assuming the SLED ratios from \cite{bothwell_2013}
(Section \ref{sec:lprimeci_lprimecotwo}). This is likely the result of
a combination of different factors, including the gas excitation properties of
individual SMGs; a redshift effect due to the evolution of the strength of the
radiation field $\langle U(z) \rangle \propto L_{\rm IR}/M_{\rm
  dust}$; the different selection techniques; and the heterogeneity of
the SMG population \citep{carilli_2013}.

\subsection{The abundance of neutral atomic carbon in galaxies}
\label{sec:ci_abundance}
\begin{figure}
  \centering
  \includegraphics[width=\columnwidth]{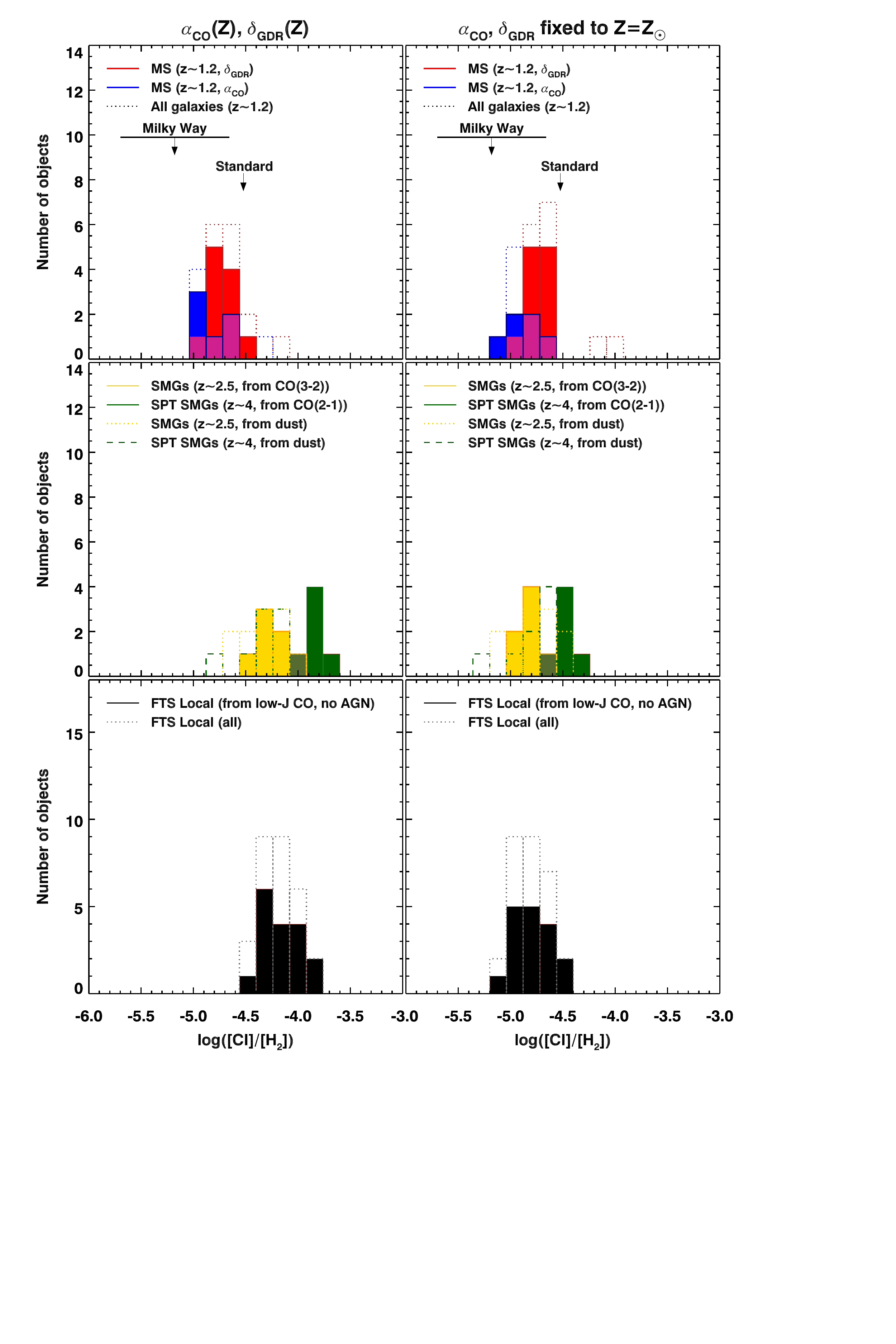}
  \caption{\textbf{Atomic carbon abundances.} \textit{Top:} The red
  and blue filled histograms show the distribution of the carbon abundances
  \ci/[H$_2$] for our sample of MS galaxies with $>3\sigma$ detections
  of \cione\ derived using a dust- and \cotwo-based $M_{\rm H_2}$
  (Section \ref{sec:h2_masses}). The black segment and arrows
    show the range of values reported in literature for regions of the Milky Way
  ($\mathrm{log}([\mathrm{C\,\scriptstyle{I}}] /
[\mathrm{H}_2]) = (-5.7, -4.7)$, \citealt{frerking_1989}), and the
  commonly adopted value of
  $\mathrm{log} ([\mathrm{C\,\scriptstyle{I}}] / [\mathrm{H}_2]) =
  -4.5$ \citep{weiss_2003}. The dotted lines mark
  the distributions including SBs and AGN, when
  possible. \textit{Center:} The yellow
  and green filled (dashed/dotted) histograms show \ci/[H$_2$] for SMGs from W11/AZ13 and
  B17 using a \cotwo- (dust)-based $M_{\rm H_2}$. \textit{Bottom:}
  The black filled histograms show \ci/[H$_2$] for
  local LIRGs without AGN signatures using \co-based $M_{\rm
    H_2}$. The dotted histogram show the abundance distribution for
  the whole FTS sample, including active galaxies. \textit{Left
    panels:} $M_{\rm H_2}$ derived adopting metallicity dependent
  $\alpha_{\rm CO}$ and $\delta_{\rm GDR}$ for MS galaxies and fixed
  $\alpha_{\rm CO}=0.8$~\msun/(\kkmspc) and $\delta_{\rm GDR}=30$ for SBs,
  local FTS galaxies, and SMGs. \textit{Right panels}: Universal
  $\alpha_{\rm CO}=3.3$~\msun/(\kkmspc) and $\delta_{\rm GDR}=85$,
  following \cite{magdis_2012} and fixing $Z=Z_\odot$ for all galaxies.}
  \label{fig:abundances}
\end{figure}

While not requiring a standard $\alpha$ or $X$ factor as
optically thick $^{12}$CO transitions, \ci\ line luminosities can be
converted into total gas masses only with prior
  knowledge of the abundance of carbon in the neutral atomic phase
  \ci/[H$_2$]. Such a conversion is necessary for any species other
than H$_2$, the dominant form of molecular gas. We derived the atomic carbon
abundances as
$[\mathrm{C\,\scriptstyle{I}}] / [\mathrm{H}_2] =
M_{\rm{[C\,\scriptscriptstyle{I}\scriptstyle{]}}} / (6\, M_{\rm{H_2}})$,
using the $M_{\rm{H_2}}$ estimates from dust and \co. Notice
that $M_{\rm{H_2}}$ does not include the helium contribution. Previous
  works adopted this or alternative approaches, providing atomic carbon abundance
  estimates in a variety of environments at different redshifts
  \citep[e.g.,][W11, AZ13 and B17 among the others]{stutzki_1997, ikeda_2002, weiss_2003,
    weiss_2005, israel_2015}. Here we redetermined the abundances
  based on a homogeneous set of assumptions, so to directly compare
  datasets in a consistent way. The discrepancies among our estimates
  and the ones in the original papers arise mainly from the choice of galaxies representative of
  the various populations (e.g., we exclude QSOs from the
  calculations), and different assumption (gas conversion
  factors, CO excitation ladder, inclusion or not of upper
  limits, dust- or CO-based gas masses, etc).
  First, we show the distribution of the estimated
abundances for the MS sample at $z\sim1.2$ in Figure \ref{fig:abundances}. The mean
values are $\mathrm{log(}[\mathrm{C\,\scriptstyle{I}}] /
[\mathrm{H}_2]) = (-4.7\pm 0.1)$ and $(-4.8\pm0.2)$ adopting $M_{\rm
  H_2}(\rm{dust})$ and $M_{\rm H_2}(\rm{CO})$,
respectively. The uncertainties represent the standard deviation
of the observed distributions for $12$ MS galaxies with
  detected continuum emission in the Rayleigh-Jeans tail and \cione\
  coverage ($11$ detections and $1$ upper limit) when using $M_{\rm
  H_2}(\rm{dust})$, and for $8$ objects with \cione\ coverage
  ($6$ detections and $2$ upper limits) and \cotwo\
detections in the case of $M_{\rm H_2}(\rm{CO})$. We excluded 1
MS source with continuum detection due to unsuccessful far-infrared
photometric deblending. We included the upper limits on
  \cione\ using a survival analysis, but their effect is negligible.
The \ciabundance\ in MS galaxies at $z\sim1.2$ is consistent with the observed values in highly obscured
clouds of the Milky Way ($\mathrm{log(}[\mathrm{C\,\scriptstyle{I}}] /
[\mathrm{H}_2]) \sim -5.7, - 4.7$
depending on the dust attenuation, \citealt{frerking_1989}, Figure
\ref{fig:abundances}), and it is lower than the typically
assumed abundance of $\mathrm{log(}[\mathrm{C\,\scriptstyle{I}}] /
[\mathrm{H}_2])=-4.5$ \citep{weiss_2003}. For reference, we also show
the distributions including SBs and the AGN-dominated objects. We
remark that the abundances presented in this work are global,
galaxy-integrated estimates, while local measurements often focus to
individual clouds. A direct comparison should be drawn with caution,
as it would be natural to find lower abundances on global scales, if
\ci\ and $\mathrm{H_2}$ are not fully cospatial. 
We similarly rederived the atomic carbon
abundances for the literature sample, using both $M_{\rm
  H_2}(\rm{CO})$ and $M_{\rm H_2}(\rm{dust})$ when
available. For the local sample of $17$ sources without AGN signatures
and \coone\ detections we find
$\mathrm{log(}[\mathrm{C\,\scriptstyle{I}}] / [\mathrm{H}_2]) = (-4.2
\pm 0.2)$; for the SPT-SMGs at $z\sim4$ from B17,
$\mathrm{log(}[\mathrm{C\,\scriptstyle{I}}] /
[\mathrm{H}_2]) = (-3.9 \pm 0.1) $ and $(-4.3\pm0.2)$ using \cotwo\ and dust,
respectively; for SMGs at $z>2.5$ from W11 and AZ13,
$\mathrm{log(}[\mathrm{C\,\scriptstyle{I}}] /
[\mathrm{H}_2]) = (-4.2\pm 0.1)$ and $(-4.3\pm 0.2)$ using \cotwo\ and
dust, respectively. The uncertainties represent the dispersion of the
distributions in Figure \ref{fig:abundances} and include upper
limits with a survival analysis. Note that $[\mathrm{C\,\scriptstyle{I}}] / [\mathrm{H}_2] =
M_{\rm{[C\,\scriptscriptstyle{I}\scriptstyle{]}}} / (6\,
M_{\rm{H_2}})$ faithfully represents the abundance of atomic carbon relative to
the molecular hydrogen $\rm{H_2}$, a good approximation for the total
gas mass $M_{\rm gas}$ at high redshift. However, both $M_{\rm
  H_2}(\rm{dust})$ and $M_{\rm H_2}(\rm{CO})$ formally include
\hi, which might be the dominant phase in local systems. Removing \hi\ and
considering the molecular gas phase only would further increase the
\ciabundance\ values reported here above for the local galaxies.

\section{The calibration of \ci\ as a gas tracer in galaxies}
\label{sec:discussion}

\subsection{The limitations of a fully empirical approach}
So far we have proven that it is feasible to detect \cione\ lines not
only in distant, extreme, and often lensed systems, such as
SMGs and QSOs, but also in normal MS galaxies at moderately
high redshifts. We showed the existence of a \lir\ -- \lprimeci\
correlation similar to the standard \lir\ -- \lprimecoone\ relation,
with \lprimeci/\lir\ ratios systematically decreasing with increasing \lir\ and distance from 
the main sequence. The strong correlation between \lprimeci\ and \lir\ makes the
latter a useful tool to predict \ci\ emission in distant galaxies.
Moreover, the roughly constant \lprimeci/\lprimecotwo\
and \mci/\mdust\ ratios on kpc-scales, regardless of total \lir, galaxy type,
and redshift, reinforce the
connection between \ci, \co, and dust, supporting the use of
\ci\ as a molecular gas tracer.\\ 

Constant \lprimeci/\lprimecotwo\ and \mci/\mdust\ ratios directly
translate into systematically lower \textit{neutral atomic}
carbon abundances in MS galaxies than in SBs/SMGs, owing to the
canonical empirical $\alpha_{\rm CO}$ and gas-to-dust $\delta_{\rm
  GDR}$ conversion factors ($\alpha_{\rm CO}\sim4 -
0.8$~\msun/(\kkmspc) and $\delta_{\rm
  GDR}\sim85 - 30$ for MS and SB galaxies,
respectively). Assuming identical conversion factors, the
  abundances are similar in MS galaxies and SBs/SMGs (we show the case
  of constant $\alpha_{\rm CO}, \delta_{\rm GDR} (Z=Z_\odot)$ in Figure \ref{fig:abundances}).
In other words, the
well-known uncertainties of the standard CO and dust tracers
affect the \textit{empirical} calibration of \ci. This practically limits
the use of this potentially superior tracer of gas in
galaxies, in absence of a calibration fully independent of the current
assumptions. Moreover, these results suggest
that the use of a universal abundance at low and
high redshift, and regardless of the galaxy population, can strongly
bias the gas masses derived from \ci\ ($M_{\rm H_2}$(\ci) scales as
(\ciabundance)$^{-1}$), as in the case of the widespread
\ciabundance$=3\times10^{-5}$ value adopted in literature, following
an estimate by \citet{weiss_2003} in a high-redshift QSO and the
average abundance reported by \citet{papadopoulos-greve_2004}.
The ascertained redshift evolution of metallicity in galaxies and the complex and history of carbon production \citep[e.g.,][]{chiappini_2003}
argue against the use of universal abundance values, even if an early and quick
enrichment might mitigate this issue in the cosmic ages explored so far. 
\begin{figure}
  \centering
  \includegraphics[width=\columnwidth]{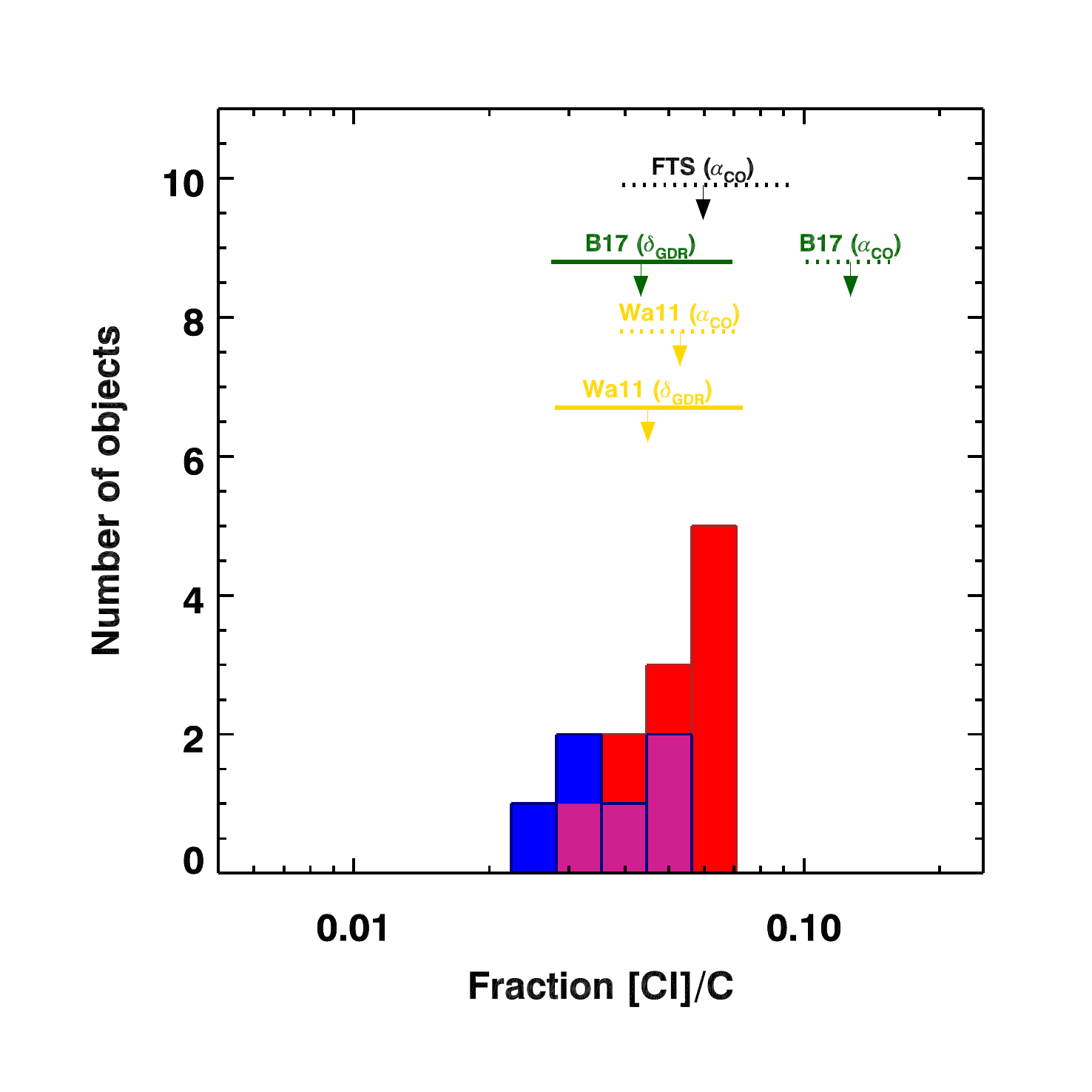}
  \caption{\textbf{Atomic carbon fractions.} The red
  and blue filled histograms show the distribution of the carbon fractions
  $f_{\rm{[C\,\scriptscriptstyle{I}\scriptstyle{]}}} =
  M_{\rm{[C\,\scriptscriptstyle{I}\scriptstyle{]}}} / M_{\rm{C}}$ for
  our sample of MS galaxies with $>3\sigma$ detections
  of \cione\ derived using a dust- and \cotwo-based $M_{\rm H_2}$,
  adopting $\delta_{\rm GDR}(Z) = \delta_{\rm GDR}(Z_\odot)$ and
  $\alpha_{\rm CO}(Z) = \alpha_{\rm CO}(Z_\odot)$. The
  colored segments and arrows show the mean \fci\ values and the
  standard deviation of the logarithmic distributions
  of SMGs and local LIRGs, as coded
  in the labels (B17: SPT-SMGs at $z\sim4$, Wa11: SMGs at $z\sim2.5$;
  FTS: Local LIRGs, no AGN).}
  \label{fig:fractions}
\end{figure}

\subsection{The small fraction of atomic carbon in galaxies}
A precise calibration of \ci\ as a total gas tracer could be intrinsically
difficult, since it requires the tracking of the small fraction of
carbon produced in galaxies in the atomic gas phase. Neglecting the
carbon locked into stars, the gas mass fraction of the neutral atomic carbon phase 
$f_{\rm{[C\,\scriptscriptstyle{I}\scriptstyle{]}}} =
M_{\rm{[C\,\scriptscriptstyle{I}\scriptstyle{]}}} / M_{\rm{C}}$ can be
derived from the definition of the mass fraction of metals:
\begin{equation} 
\label{eq:metal}
Z = \frac{M_{\rm metals}}{M_{\rm H_2}} =
\frac{M_{\rm metals}}{M_{\rm
    C}}\frac{M_{\rm{C}}}{M_{\rm{[C\,\scriptscriptstyle{I}\scriptstyle{]}}}}
\frac{M_{\rm{[C\,\scriptscriptstyle{I}\scriptstyle{]}}}}{M_{\rm H_2}}
\end{equation}
where $M_{\rm C}$ is the total mass of carbon, \mci\
the mass in the neutral atomic phase, $M_{\rm metals}$ the total mass
of metals, and $M_{\rm H_2}$ is the hydrogen gas mass, excluding helium. 
From Eq. \ref{eq:metal}, we derive $f_{\rm{[C\,\scriptscriptstyle{I}\scriptstyle{]}}} \propto
\frac {M_{\rm{[C\,\scriptscriptstyle{I}\scriptstyle{]}}}}{M_{\rm
    dust}\delta_{\rm GDR}/\mathrm{He}}$ or $\propto
\frac {M_{\rm{[C\,\scriptscriptstyle{I}\scriptstyle{]}}}}{L'_{\rm
    CO}\alpha_{\rm CO}/\mathrm{He}}$. Assuming a solar metallicity and composition
($Z_\odot=0.0134$ and $\frac{M_{\rm
    C}}{M_{\rm metals}}=0.1779$, \citealt{asplund_2009}), the mass fraction of carbon in the atomic
gas phase in the MS galaxies is $4.6^{+1.3}_{-1.0}$\% and
$3.4^{+1.4}_{-1.0}$\% using \mdust\ and \lprimecotwo, respectively
(Figure \ref{fig:fractions}). These estimates and their
  uncertainties include upper limits through survival analysis and
  they represent the mean and standard deviation of the logarithmic distributions in
Figure \ref{fig:fractions}. Note that these values depend on the
choice of metallicity and carbon fraction adopted here.
Using the metallicity from the Fundamental Metallicity Relation does not
impact this result (Section \ref{sec:h2_masses}). We derive similar
$f_{\rm{[C\,\scriptscriptstyle{I}\scriptstyle{]}}} =
6.0^{+3.3}_{-2.1}$\% for LIRGs; $4.5^{+2.8}_{-1.7}$\% and
$5.3^{+1.9}_{-1.4}$\% for SMGs at $z\sim2.5$ from W11/AZ13 using dust
and \co\, respectively; $4.3^{+2.6}_{-1.6}$\% and
$12.7^{+3.2}_{-2.6}$\% for SPT-SMGs from B17 adopting the dust- and
\co-based calibration and including upper limits. We used dust and CO
to derive \mgas, and assuming a metallicity of $Z\sim2.8\,Z_\odot$
corresponding to $\alpha_{\rm CO} = 0.8$~\msun/(\kkmspc) and $\delta_{\rm
  GDR}=30$. Supersolar metallicities are necessary
to obtain these commonly adopted values when using the parametrization
by \citet[their Eq. 8 with the corrected intercept discussed above and Section 4.2]{magdis_2012}.
However, while this likely has strong physical roots for the optically thin dust emission, 
$\alpha_{\rm CO}$ values of $\sim0.8$~\msun/(\kkmspc) could be found also at $Z\sim Z_\odot$, being
the CO emission optically thick and, thus, critically dependent on
other parameters (i.e., the FWHM of the line).\\

The atomic carbon fractions \fci\ in Figure \ref{fig:fractions}
reflect the similar \lprimeci/\lprimecotwo\ and \mci/\mdust\ ratios
for MS and SBs/SMGs in Figure \ref{fig:ci_co_dust}, analogously to the
\ci\ abundances shown in Figure \ref{fig:abundances}. From a theoretical perspective, even simple plan-parallel
PDR models could partially explain the observed constant \cione/\cotwo\ ratio by the relative
insensitivity of the \ci\ emission to the strength of UV radiation. In
fact, while a stronger radiation field pushes
the C-to-CO transition deeper into the gas slab, the size of the \ci-
and \cotwo-emitting regions stays relatively constant \citep[e.g.,][and many others]{kaufman_1999}.
Our $f_{\rm{[C\,\scriptscriptstyle{I}\scriptstyle{]}}}$ estimates suggest that \ci\ represents a very minor fraction of the
overall mass of carbon in galaxies, as the majority is 
in CO molecules ($N($\ci$)/N($CO$)=0.1-0.2$, \citealt{ikeda_2002}) and
depleted on dust ($\sim27$\% of the overall carbon abundance, e.g.,
\citealt{vandishoeck_1988}), neglecting the quantity locked in stars.  
The small mass fraction of atomic carbon and the associated low
  column densities explain the small optical depth of the \cione\ line, a major
  advantage in the use of this species to trace the gas content of
  galaxies \citep{papadopoulos_2004}. However, an accurate assessment
  of such minimal \ci\
  fractions and the detection of
  relative variations in different galaxy populations -- if present --
  are complicated by both observational and theoretical uncertainties
  (e.g., the history of chemical enrichment in galaxies, 
  affecting any tracers of the molecular
  gas mass other than H$_2$).

\subsection{Possible hidden systematic variations of \ciabundance}
We showed that MS and SB/SMGs have similar
  \lprimecifull/\lprimecotwo\ and \mci/\mdust\ ratios (Figure
\ref{fig:ci_co_dust}), resulting in different \ciabundance\ owing to
standard assumptions on the dust/CO-to-gas conversion
factors. However, reversing the argument, \textit{intrinsic} large variations of \ciabundance\
might \textit{not} directly translate into large differences in the
observed ratios, due to the counter-effect of higher
$\delta_{\rm GDR}$ and $\alpha_{\rm CO}$ for MS galaxies than
for SBs. Analogously, intrinsic systematic differences of \ciabundance\ between
SB and MS would blur large variations of SFE into similar
\lir/\lprimeci\ ratios, equivalently to what is observed for CO
\citep{daddi_2010b}. The opposite dependence
of \ciabundance\ and $\delta_{\rm GDR}$ on metallicity could explain the
small variations of \mci/\mdust\
observed in our data compilation: metal rich SB
galaxies tend to have larger \ciabundance\ abundances than MS objects,
compensating lower gas-to-dust ratios $\delta_{\rm  GDR}$. 
On the other hand, parameters other than metallicity should play a major role
in the comparison of optically thick (CO) and thin (\ci, dust) tracers
\citep[e.g., turbulent velocities and compression resulting in broad
CO lines,][]{bournaud_2015}. On top of these effects, enhanced cosmic
ray rates in sources with large SFEs \citep[e.g.,][]{papadopoulos_2004, bisbas_2015, bisbas_2017} could increase
\ciabundance, further reducing strong variations of the observed
\mci/\mdust\ and \lprimeci/\lprimecotwo\ ratios.
The degeneracies listed here would be broken by an estimate of the gas
mass independent of the assumptions we have to
make for the currently available data.

\section{Conclusions}
\label{sec:conclusions}
We presented the first results of a survey targeting the \cione\
transition in main-sequence galaxies at $z\sim1.2$, expanding the
samples of starbursts and SMGs
currently present in literature towards
a population of normal galaxies.\\ 

We showed the existence of a
sublinear correlation between the total log(\lir) and
log(\lprimeci), with a decreasing \lprimeci/\lir\ ratio with
increasing \lir. The slope of this relation and the
$\sim4\sigma$ significant displacement of the \cione-detected MS galaxies from the
sequence traced by SBs/SMGs are similar to what is observed for
\coone. The difference will be easily tested by populating the 
distance from the main-sequence -- \lprimeci/\lir\ plane. Based on
current evidence, starbursts may have lower values of this ratio than
do main-sequence galaxies.
These observations strengthen the \ci\ -- CO relation on global
galaxy scales and suggest that \lprimeci/\lir\ is an effective tracer
of the depletion timescales in galaxies, decreasing with increasing
\lir\ and shorter in SBs/SMGs than in MS galaxies.\\

 We further showed
that the \lprimeci/\lprimecotwo\ and \mci/\mdust\ ratios in local
LIRGs, MS galaxies and SBs at $z\sim1.2$, and lensed SMGs 
at $z>2.5-4$ are similar, with an intrinsic scatter of $\sim0.2$~dex. 
These ratios are proportional to a mass fraction of carbon
in the neutral atomic phase of $f_{\rm{[C\,\scriptscriptstyle{I}\scriptstyle{]}}}
= M_{\rm{[C\,\scriptscriptstyle{I}\scriptstyle{]}}} / M_{\rm{C}}\sim
3-13$\%, roughly independent of the galaxy type and redshift, although
affected by substantial scatter. The relative
insensitivity of \ci\ emission to the strength of the radiation field
could partially explain the constant \cione/\cotwo\ ratios, as the
size of the \ci- and \co-emitting regions stays relatively constant in
different environments.\\
 
We then estimated the atomic carbon abundances -- necessary to derive the total gas mass from
\ci\ observations -- by comparing \cione, dust, and CO low-$J$ emissions.
Adopting standard $\alpha_{\rm CO}$ and gas-to-dust $\delta_{\rm GDR}$
conversion factors, we find \ciabundance~$\sim1.6-1.9\times10^{-5}$ for
MS galaxies at $z\sim1.2$. These values are $\sim 3-8\times$ lower than the abundance in
high-redshift SMGs, and $\sim4\times$ lower than in local
LIRGs. At this stage, this difference is mainly a consequence of the choice
of $\alpha_{\rm CO}$ and $\delta_{\rm GDR}$, and it currently exacerbates possible
minimal differences in the observables, if present. However, systematically
higher \ciabundance\ in SB than in MS galaxies, possibly driven by larger
metallicities and/or cosmic rays rates, might result in similar
observables, a degeneracy that cannot be
broken with the available data and tools.
All things considered, our findings caution against the use of a universal
atomic carbon abundance regardless of the galaxy type and redshift.\\

\section*{Acknowledgements}
We acknowledge the constructive comments from the anonymous referee,
which significantly improved the content and presentation of the
results. We thank Mark Sargent for providing the initial catalogs and discussions
during the early stages of this work.
FV and GEM acknowledge the Villum
Fonden research grant 13160 ``Gas to stars, stars to
dust: tracing star formation across cosmic time'' and the Cosmic Dawn Center
of Excellence funded by the Danish National Research Foundation.
YG acknowledges partial support 
of China grant no. 2017YFA0402704, NSFC grant no. 11420101002. This paper makes
use of the following ALMA data: ADS/JAO.ALMA\#2016.1.01040.S,
\#2015.1.00260.S, \#2016.1.00171.S. ALMA is a
partnership of ESO (representing its member states), NSF (USA) and
NINS (Japan), together with NRC (Canada), MOST and ASIAA (Taiwan), and
KASI (Republic of Korea), in cooperation with the Republic of
Chile. The Joint ALMA Observatory is operated by ESO, AUI/NRAO and
NAOJ. In this work we made use of the COSMOS master spectroscopic
catalog, available within the collaboration and kept updated by Mara Salvato.

\bibliography{../bib_ci_valentino}

\appendix
\section{Galaxy spectral energy distributions and spectra}
We show in Figure \ref{fig:data} the 1D spectra of our sample of $21$ galaxies 
at $z\sim1.2$ observed with ALMA Band 6. The black line show the
spectrum; the yellow area marks the channels used to measure
  the line fluxes, matching brighter \co\ lines, if present (Section
  \ref{sec:observations}, Table \ref{tab:fir}); the solid red line the best fitting Gaussian obtained fixing the 
redshift and width of the \cione\ line to match the brighter \co\
emissions and the continuum level to the
estimate over the full 7.5~GHz band; 
the dashed dark red line shows the best \cione\ line model with free
parameters for the line and a local continuum emission estimate within
$\pm2$~GHz from the line center.
When a line is not detected, we show in red the location of the expected \cione\ emission 
based on detected \co\ lines. The black thick shows the expected
position of the line based on the optical/near-infrared spectroscopic
redshift. Figure \ref{fig:data} further shows the near-infrared to 
radio photometry for our sample. The filled red circles mark the data points we
considered for our modeling (black line). The empty red circles were not considered in the
fit \citep{magdis_2012}. Arrows mark $3\sigma$ upper
limits. The solid blue line shows the best fit template for
  the dusty torus component.

\setcounter{figure}{0} \renewcommand{\thefigure}{A.\arabic{figure}}
\begin{figure*}
  \centering
  \includegraphics[angle=90, width=\textwidth]{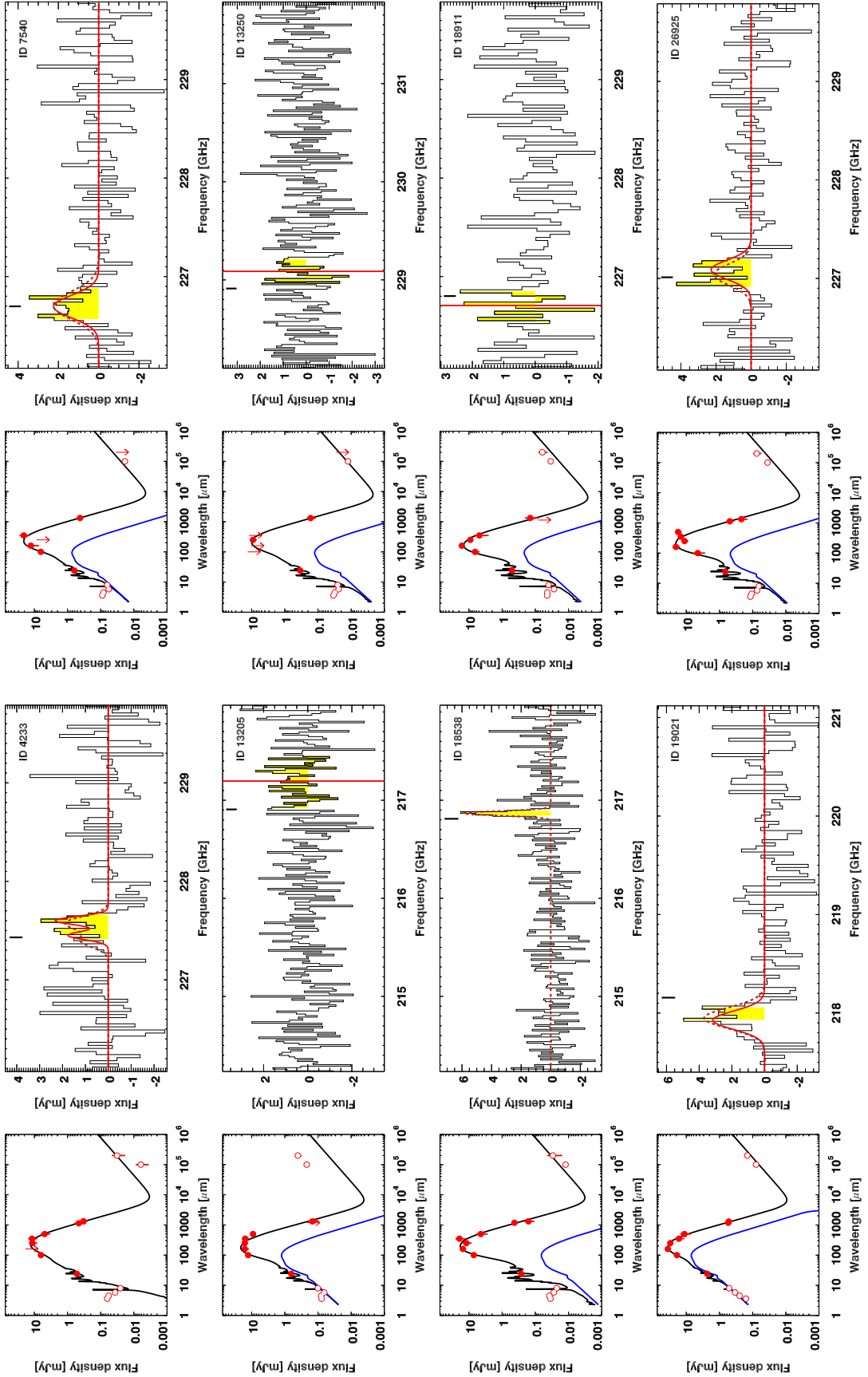}
  \caption{Far infrared SEDs and \cione\ spectra of our sample
      of $21$ galaxies at $z\sim1.2$ followed up with ALMA.}
    \label{fig:data}
\end{figure*}

\setcounter{figure}{0}    
\begin{figure*}
  \centering
  \includegraphics[angle=90, width=\textwidth]{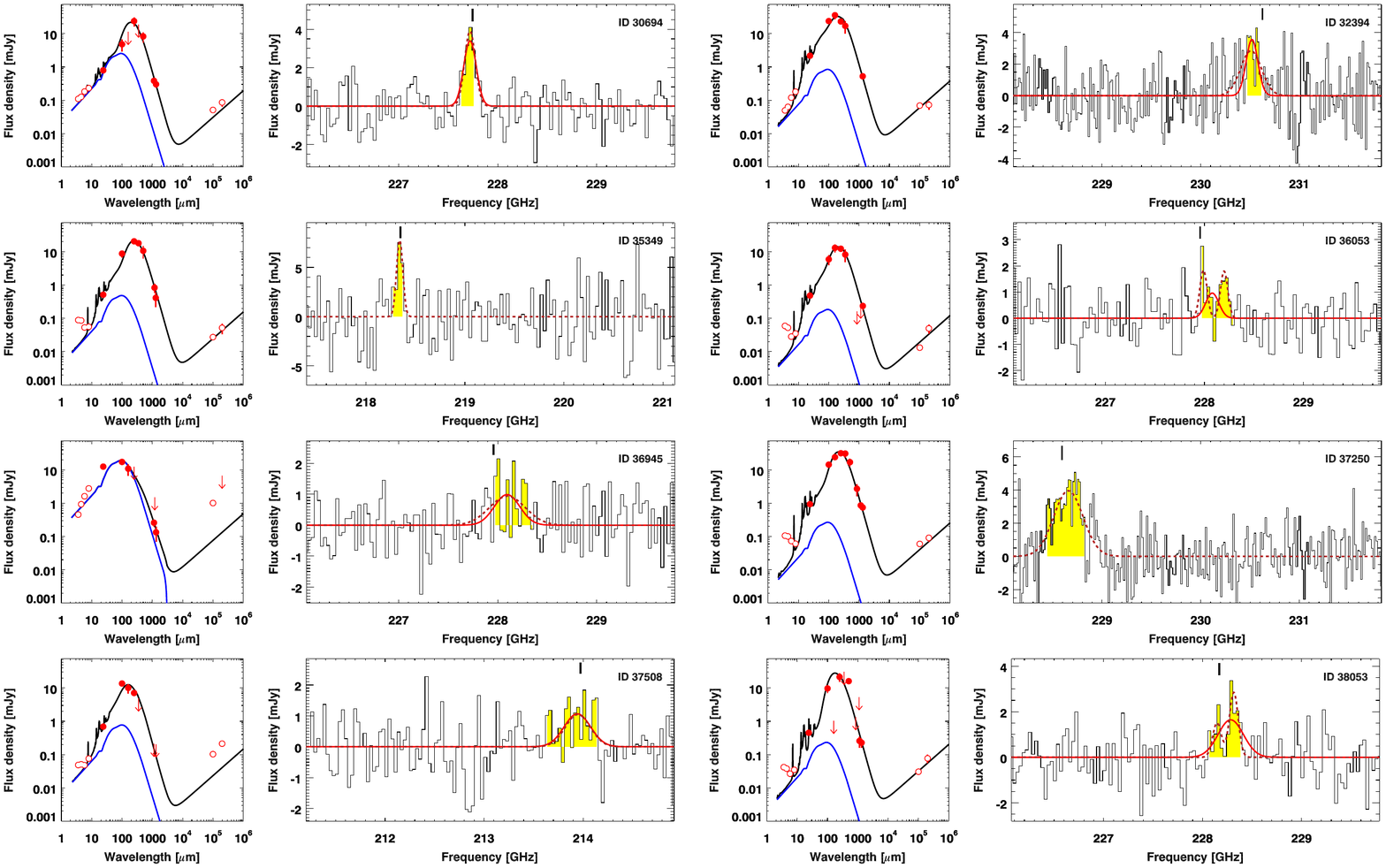}
  \caption{(continue)}
\end{figure*}

\setcounter{figure}{0}    
\begin{figure*}
  \centering
  \includegraphics[angle=90, width=\textwidth]{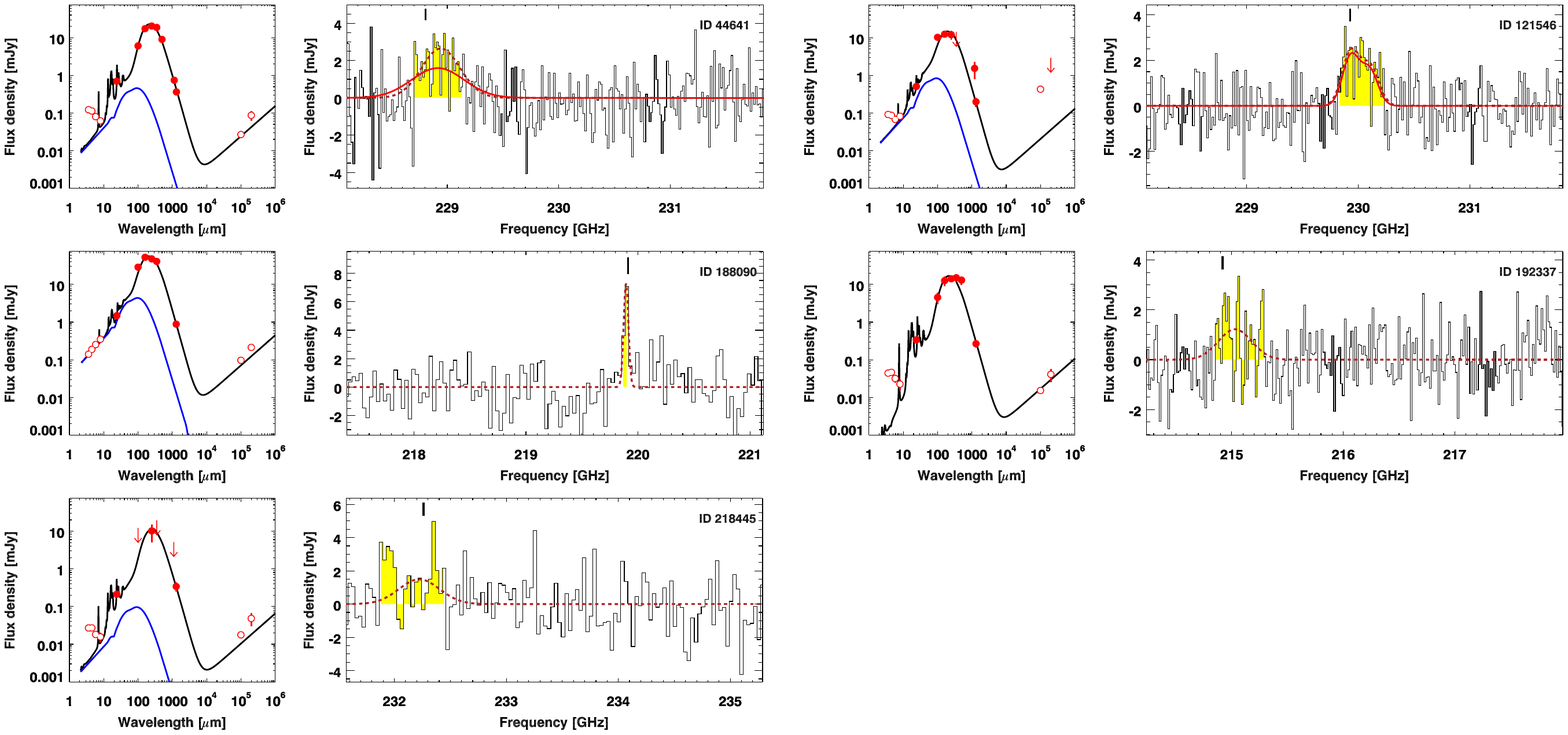}
  \caption{(continue)}
\end{figure*}


 \begin{deluxetable}{ccccccccccccc}
   \tabletypesize{\scriptsize}
   \tablecolumns{13}
   \tablecaption{Main-sequence galaxies at $z\sim1.2$\label{tab:fir}.}
   \smallskip
   \tablehead{
     \colhead{ID} & 
     \colhead{RA}&
     \colhead{Decl}&
     \colhead{$z_{\rm spec}$} & 
     \colhead{$\mathrm{log}$(\mstar)}&
     \colhead{\lir}&
     \colhead{\lfir}&
     \colhead{\lprimeci}&
     \colhead{$L'_{\rm CO(2-1)}$}&
     \colhead{\ici} & 
     \colhead{$I_{\rm{CO(2-1)}}$} &
     \colhead{$f_{\rm AGN}$} &
     \colhead{Type} \tiny\\
     \colhead{}&
     \colhead{(deg)}& 	
     \colhead{(deg)}&
     \colhead{}&
     \colhead{$M_{\odot}$}&
     \colhead{$10^{11}\,L_{\odot}$}&
     \colhead{$10^{11}\,L_{\odot}$}&
     \colhead{$10^{10}\,$~\kkmspc}&
     \colhead{$10^{10}\,$~\kkmspc}&
     \colhead{\jykms}& 	
     \colhead{\jykms}&
     \colhead{}&
     \colhead{}\\
     \colhead{(1)}&
     \colhead{(2)}& 	
     \colhead{(3)}&
     \colhead{(4)}&
     \colhead{(5)}&
     \colhead{(6)}&
     \colhead{(7)}&
     \colhead{(8)}&
     \colhead{(9)}&
     \colhead{(10)}&
     \colhead{(11)}&
     \colhead{(12)}&
     \colhead{(13)}\\
   }
   \rotate
   \startdata
   4233   	& 150.39732 &	2.24088 &	$1.1630 \pm 0.0003$\tablenotemark{a}&	10.89 &	$8.09	\pm 0.49$ &	$5.05\pm0.30$& $0.24	\pm 0.05$ & 	$-$                         &	$0.6	\pm 0.14$ &	$-$               &	$-$ &	MS  \\
   7540	& 150.41281 &	1.85157 &	$1.1714 \pm 0.0003$\tablenotemark{a}&	11.06 &	$9.36	\pm 1.32$ &	$7.36\pm1.04$& $0.34	\pm 0.08$ &	$-$                         &	$0.86	\pm 0.19$ &	$-$               &	0.12 &	MS  \\
   13205	& 150.57484 &	1.94786 &	$1.2660 \pm 0.0004$\tablenotemark{b}&	11.10 &	$12.28	\pm 1.27$ &	$9.68\pm1.00$& $<0.2	               $ &	$1.93	\pm 0.39$ &	$<0.44	        $ &	$0.92	\pm 0.18 $ &	0.16 &	MS  \\
   13250	& 150.17650 &	1.95523 &	$1.1484 \pm 0.0002$\tablenotemark{a}&	10.41 &	$5.22	\pm 1.54$ &	$3.68\pm1.09$& $<0.1	               $ &	$-$                          &	$<0.26	        $ &	$-$               &	0.04 &	MS  \\
   18538	& 150.12639 &	2.43285 &	$1.2694 \pm 0.0001$\tablenotemark{c}&	10.56 &	$9.83	\pm 9.45$ &	$8.22\pm7.91$& $0.19	\pm 0.02$ &	$-$                          &	$0.41	\pm 0.05$ &	$-$               &	0.01 &	MS  \\
   18911	& 150.04820 &	2.46767 &	$1.1709 \pm 0.0003$\tablenotemark{a}&	10.64 &	$7.78	\pm 1.68$ &	$5.87\pm1.27$& $<0.16	               $ &	$0.58	\pm 0.13$ &	$<0.39	        $ &	$0.32	\pm 0.07 $ &	0.04 &	MS  \\
   19021	& 150.05878 &	2.47739 &	$1.2581 \pm 0.0003$\tablenotemark{d}&	11.78 &	$19.27	\pm 1.57$ &	$14.57\pm1.19$& $0.35	\pm 0.07$ &	$1.90	\pm 0.24$ &	$0.76	\pm 0.14$ &	$0.91	\pm 0.11 $ &	0.41 &	AGN \\
   26925	& 150.46350 &	1.88331 &	$1.1671 \pm 0.0003$\tablenotemark{a}&	11.05 &	$9.13	\pm 1.66$ &	$7.56\pm1.37$& $0.35	\pm 0.06$ &	$1.47	\pm 0.22$ &	$0.88	\pm 0.16$ &	$0.81	\pm 0.12 $ &	0.08 &	MS  \\
   30694	& 149.66026 &	2.41099 &	$1.1606 \pm 0.0002$\tablenotemark{a}&	10.78 &	$8.67	\pm 1.45$ &	$7.58\pm1.26$& $0.24	\pm 0.03$ &	$1.62	\pm 0.21$ &	$0.61	\pm 0.09$ &	$0.91	\pm 0.12 $ &	0.32 &	MS  \\
   32394	& 150.06781 &	1.85108 &	$1.1345 \pm 0.0001$\tablenotemark{a}&	10.03 &	$22.54	\pm 6.37$ &	$12.90\pm3.65$& $0.24	\pm 0.05$ &	$-$                          &	$0.64	\pm 0.12$ &	$-$               &	0.06 &	SB  \\
   35349	& 150.49902 &	1.72446 &	$1.2543 \pm 0.0002$\tablenotemark{c}&	11.19 &	$11.5	\pm 1.83$ &	$8.65\pm1.38$& $0.41	\pm 0.11$ &	$3.57	\pm 0.44$ &	$0.89	\pm 0.25$ &	$1.72	\pm 0.21 $ &	0.07 &	MS  \\
   36053	& 149.68564 &	1.91986 &	$1.1573 \pm 0.0003$\tablenotemark{a}&	10.87 &	$7.69	\pm 3.24$ &	$5.38\pm2.26$& $0.17	\pm 0.05$ &	$-$                         &	$0.43	\pm 0.13$ &	$-$               &	0.04 &	MS  \\
   36945	& 150.19550 &	2.12404 &	$1.1569 \pm 0.0003$\tablenotemark{a}&	11.43 &	$0.44	\pm 0.03$ &	$0.37\pm0.02$& $0.17	\pm 0.05$ &	$1.31	\pm 0.37$ &	$0.44	\pm 0.14$ &	$0.47	\pm 0.13 $ &	0.99 &	AGN \\
   37250	& 149.61813 &	2.19346 &	$1.1526 \pm 0.0002$\tablenotemark{c}&	10.96 &	$16.21	\pm 6.97$ &	$12.48\pm5.37$& $0.68	\pm 0.06$ &	$4.58	\pm 0.32$ &	$1.76	\pm 0.16$ &	$2.6	\pm 0.18 $ &	0.03 &	MS  \\
   37508	& 150.38953 &	2.25585 &	$1.3020 \pm 0.0003$\tablenotemark{b}&	11.04 &	$11.20	\pm 2.17$ &	$6.38\pm1.24$& $0.31	\pm 0.08$ &	$1.48	\pm 0.42$ &	$0.64	\pm 0.16$ &	$0.43	\pm 0.12 $ &	0.12 &	MS  \\
   38053	& 150.55685 &	2.39140 &	$1.1562 \pm 0.0003$\tablenotemark{a}&	10.66 &	$12.90	\pm 4.03$ &	$10.84\pm3.38$& $0.26	\pm 0.05$ &	$1.69	\pm 0.31$ &	$0.67	\pm 0.14$ &	$0.96	\pm 0.17 $ &	0.03 &	SB  \\
   44641	& 150.65852 &	2.31786 &	$1.1495 \pm 0.0002$\tablenotemark{a}&	11.10 &	$9.19	\pm 1.99$ &	$7.72\pm1.67$& $0.44	\pm 0.06$ &	$1.63	\pm 0.35$ &	$1.16	\pm 0.17$ &	$0.93	\pm 0.20  $ &	0.07 &	MS  \\
   121546$^\dagger$& 150.05910 &	2.21995 &	$1.1392 \pm 0.0002$\tablenotemark{a}&	11.08 &	$6.89	\pm 0.94$ &	$5.28\pm0.72$& $0.39	\pm 0.04$ &	$-$                          &	$1.04	\pm 0.12$ &	$-$               &	0.16 &	MS  \\
   188090& 150.42799 &	2.55860 &	$1.2383 \pm 0.0002$\tablenotemark{c}&	11.12 &	$28.97	\pm 1.22$ &	        $22.74\pm0.96$& $0.23	\pm 0.04$ &	$-$                          &	$0.51	\pm 0.09$ &	$-$               &	0.22 &	SB  \\
   192337& 150.42073 &	2.62297 &	$1.2884 \pm 0.0003$\tablenotemark{c}&	10.90 &	$9.06	\pm 0.64$ &	        $7.50\pm0.53$& $0.30	\pm 0.07$ &	$-$                          &	$0.62	\pm 0.14$ &	$-$               &	$-$ &	MS  \\
   218445& 149.92007 &	2.61905 &	$1.1199 \pm 0.0004$\tablenotemark{c}&	10.60 &	$3.83	\pm 2.19$ &	        $3.25\pm1.86$& $0.49	\pm 0.09$ &	$-$                          &	$1.33	\pm 0.25$ &	$-$               &	0.04 &	MS  \\
  \enddata
   \tablecomments{\scriptsize 1 - ID; 2, 3 - RA, Decl J2000; 4 -
     Spectroscopic redshift from sub-millimeter emission lines:
     \textit{a:} From \cofour; \textit{b:} \cofive; \textit{c:}
     \cione; \textit{d:} \cotwo; 5 -  Stellar mass from SED modeling
     \citep{muzzin_2013, laigle_2016}. Uncertainty: 0.2~dex; 6 - Total
     infrared luminosity integrated within $8-1000$~$\mu$m due to star
     formation (i.e., corrected for torus emission); {7 - Far-infrared luminosity 
     integrated within $40-400$~$\mu$m due to star
     formation (i.e., corrected for torus emission); 8, 9 -
     Galaxy-integrated \lprimeci\ and \lprimecotwo; 10, 11 - Galaxy
     velocity integrated \cione\ and \cotwo\ fluxes. These estimates
     are obtained as weighted average flux densities within the
     channels maximizing the SNR of the lines, times the velocity width covered by
     these channels, and a 10\% correction to total fluxes from the
     comparison with the best Gaussian modeling with all free parameters;}
   12 - Fraction of infrared AGN emission: $L_{\rm AGN+SF} = L_{\rm IR} /
   (1-f_{\rm AGN})$; 
   13 - Galaxy type: MS = main-sequence; SB = Starburst
   ($>3.5\times$ above the main-sequence); AGN = SED dominated by
   torus emission.\\ 
\\
$^\dagger$: selected based on $z_{\rm opt/NIR}$ from \cite{casey_2012b}.\\
Upper limits are at $3\sigma$.}
 \end{deluxetable}

\newpage

\begin{deluxetable}{ccccccc}
    \tabletypesize{\normalsize}
      \tablecolumns{7}
	\tablecaption{Main-sequence galaxies at $z\sim1.2$ (continue). \label{tab:mgas}}
       \smallskip
       \tablehead{
         \colhead{ID} & 
         \colhead{\mdust}&
         \colhead{$\delta_{\rm GDR}$}&
         \colhead{$\alpha_{\rm CO}$}&
         \colhead{\mci} & 
         \colhead{\ciabundance$_{\rm(dust)}$}&
         \colhead{\ciabundance$_{\rm(CO)}$} \\
         \colhead{}&
         \colhead{$10^{8}M_{\odot}$}& 	
         \colhead{}&
         \colhead{}&
         \colhead{$10^{6}M_{\odot}$}&
         \colhead{$\times 10^{-5}$}&
         \colhead{$\times 10^{-5}$}\\
         \colhead{(1)}&
         \colhead{(2)}& 	
         \colhead{(3)}&
         \colhead{(4)}&
         \colhead{(5)}&
         \colhead{(6)}&
         \colhead{(7)}\\
       }
       \startdata
           4233&   $5.8\pm1.1$&   $82.0$&   $-$&   $3.0\pm0.7$&   $1.4\pm0.8$&   $-$ \\
           7540&   $7.7\pm1.2$&   $72.5$&   $-$&   $4.3\pm0.9$&   $1.7\pm0.9$&   $-$ \\
          13205&   $2.3\pm0.3^*$&   $72.3$&   $2.6$&   $<2.5$&   $<3.4$&   $<1.0$ \\
          13250&   $2.6\pm0.7$&   $121.5$&   $-$&   $<1.3$&   $<0.9$&   $-$ \\
          18538&   $3.5\pm0.5$&   $114.3$&   $-$&   $2.4\pm0.3$&   $1.4\pm0.7$&   $-$ \\
          18911&   $1.2\pm0.2^*$&   $102.5$&   $4.2$&   $<1.9$&   $<3.7$&   $<1.5$ \\
          19021&   $5.7\pm0.5$&   $76.3$&   $2.8$&   $4.3\pm0.8$&   $2.3\pm1.1$&   $1.6\pm0.8$ \\
          26925&   $4.6\pm0.7$&   $72.8$&   $2.6$&   $4.4\pm0.8$&   $2.9\pm1.5$&   $2.2\pm1.1$ \\
          30694&   $4.2\pm0.4$&   $91.2$&   $3.6$&   $3.0\pm0.4$&   $1.8\pm0.9$&   $1.0\pm0.5$ \\
          32394&   $7.6\pm1.6$&   $30.0$&   $-$&   $3.0\pm0.6$&   $3.0\pm1.6$&   $-$ \\
          35349&   $12.0\pm2.8$&   $76.3$&   $2.8$&   $5.1\pm1.4$&   $1.3\pm0.7$&   $1.0\pm0.5$ \\
          36053&   $3.2\pm0.7$&   $82.9$&   $-$&   $2.1\pm0.6$&   $1.8\pm1.1$&   $-$ \\
          36945&   $4.0\pm0.7$&   $76.3$&   $2.8$&   $2.2\pm0.7$&   $1.6\pm0.9$&   $1.1\pm0.7$ \\
          37250&   $9.6\pm0.8$&   $83.9$&   $3.2$&   $8.5\pm0.8$&   $2.4\pm1.1$&   $1.1\pm0.5$ \\
          37508&   $0.9\pm0.1^*$&   $75.1$&   $2.7$&   $3.9\pm1.0$&   $12.6\pm6.6$&   $1.8\pm1.1$ \\
          38053&   $3.1\pm0.4$&   $30.0$&   $0.8$&   $3.3\pm0.7$&   $8.0\pm4.2$&   $4.6\pm2.5$ \\
          44641&   $7.0\pm0.7$&   $70.2$&   $2.5$&   $5.5\pm0.8$&   $2.6\pm1.3$&   $2.6\pm1.4$ \\
         121546&   $2.7\pm0.7^{\dagger}$&   $69.1$&   $-$&   $4.9\pm0.6$&   $6.0\pm3.2$&   $-$ \\
         188090&   $8.4\pm0.8$&   $30.0$&   $-$&   $2.8\pm0.5$&   $2.6\pm1.3$&   $-$ \\
         192337&   $4.6\pm0.6$&   $82.4$&   $-$&   $3.7\pm0.8$&   $2.2\pm1.2$&   $-$ \\
         218445&   $6.4\pm1.7$&   $97.0$&   $-$&   $6.1\pm1.1$&   $2.2\pm1.2$&   $-$ \\
\enddata
\tablenotemark{$^*$}{\footnotesize No continuum detection in the Rayleigh-Jeans tail of the dust emission. This value should be treated as an order of magnitude estimate.}\\
\tablenotemark{$^{\dagger}$}{\footnotesize Significant blending of the
  photometry.}
\tablecomments{\footnotesize 1 - ID; 2 - Dust mass from SED modeling;
  3 - Gas-to-dust conversion factor. Value fixed to $\delta_{\rm
    GDR}=30$ for starbursts. Uncertainty: 0.2~dex; 4 - $\alpha_{\rm
    CO}$ conversion factor. Value fixed to $\alpha_{\rm
    CO}=0.8$~\msun/(\kkmspc) for starbursts. Uncertainty: 0.2~dex; 5 -
  Mass of \ci\ (from \lprimeci\ assuming $T_{\rm ex} = 30$~K); 6 -
  Atomic carbon abundance adopting \mgas(dust)  and removing the $1.36\times$ contribution
  of helium; 7 - Atomic carbon
  abundance adopting \mgas(\co) and removing the $1.36\times$ contribution
  of helium.\\   
  \\
  Upper limits are $<3\sigma$. Systematic uncertainties on \mgas\ are
  not included in the error budget.}
\end{deluxetable}

 \begin{deluxetable}{lccccccccccccc}
   \tabletypesize{\tiny}
   \tablecolumns{13}
   \tablewidth{0pt}
   \tablecaption{Local galaxies.\label{tab:firlocal}}
   \smallskip
   \tablehead{
     \colhead{ID} & 
     \colhead{$D_{\rm L}$}&
     \colhead{$z_{\rm spec}$}&
     \colhead{$\mathrm{log}$(\mstar)}&
     \colhead{\lir}&
     \colhead{\lprimeci}&
     \colhead{$L'_{\rm CO(1-0)}$}&
     \colhead{$L'_{\rm CO(2-1)}$}&
     \colhead{\ici}&
     \colhead{$I_{\rm CO(1-0)}$}&
     \colhead{$I_{\rm CO(2-1)}$}&
     \colhead{$\eta_{\rm \,beam}$}&
     \colhead{AGN} \tiny\\
     \colhead{}&
     \colhead{Mpc}& 	
     \colhead{}&
     \colhead{$M_{\odot}$}&
     \colhead{$10^{11}\,L_{\odot}$}&
     \colhead{$10^{8}\,$~\kkmspc}&
     \colhead{$10^{8}\,$~\kkmspc}&
     \colhead{$10^{8}\,$~\kkmspc}&
     \colhead{$10^{2}\,$~\jykms}&
     \colhead{$10^{2}\,$~\jykms}&
     \colhead{$10^{2}\,$~\jykms}&
     \colhead{}&
     \colhead{}\\
     \colhead{(1)}&
     \colhead{(2)}& 	
     \colhead{(3)}&
     \colhead{(4)}&
     \colhead{(5)}&
     \colhead{(6)}&
     \colhead{(7)}&
     \colhead{(8)}&
     \colhead{(9)}&
     \colhead{(10)}&
     \colhead{(11)}&
     \colhead{(12)}&
     \colhead{(13)}\\
   }
   \rotate
   \startdata
   Arp193&             101.6&           0.02330&              10.5&   $4.8\pm1.0$ &   $10.4\pm3.3$&   $39.4\pm4.2$&   $35.3\pm5.5$&   $7.7\pm2.4$&   $1.6\pm0.2$&   $5.7\pm0.9$&   $1$&   N \\
   Arp220&              78.7&           0.01813&              10.7&   $16.4\pm3.3$ &   $13.8\pm2.8$&   $81.0\pm4.5$&   $77.9\pm15.7$&   $16.9\pm3.4$&   $5.4\pm0.3$&   $20.9\pm4.2$&   $1.21$&   Y \\
   Arp299-A&              44.9&           0.01041&         $-$&   $2.9\pm0.6$ &   $4.8\pm1.2$&   $28.4\pm3.2$&   $30.1\pm2.4$&   $17.8\pm4.3$&   $5.8\pm0.7$&   $24.6\pm1.9$&   $1.37$&   N \\
   Arp299-B&              44.9&           0.01041&         $-$&   $1.6\pm0.3$ &   $7.3\pm2.1$&   $18.8\pm2.2$&   $26.7\pm2.5$&   $27.2\pm7.9$&   $3.8\pm0.4$&   $21.8\pm2.1$&   $1.29$&   N \\
   Arp299-C&              44.9&           0.01041&         $-$&   $1.4\pm0.3$ &   $6.8\pm1.6$&   $17.7\pm2.1$&   $21.8\pm2.2$&   $25.2\pm6.0$&   $3.6\pm0.4$&   $17.8\pm1.8$&   $1.26$&   N \\
   CGCG049-057&              56.2&           0.01300&               9.8&   $1.8\pm0.4$ &   $2.3\pm0.7$&   $-$&   $-$&   $5.5\pm1.6$&   $-$&   $-$&   $-$&   N \\
   ESO173-G015&              42.0&           0.00974&              10.4&   $3.7\pm0.7$ &   $7.1\pm1.5$&   $-$&   $-$&   $30.1\pm6.4$&   $-$&   $-$&   $1.88$&   N \\
   IRASF18293-3413&              78.9&           0.01818&              10.9&   $6.1\pm1.2$ &   $20.9\pm2.6$&   $118.6\pm11.0$&   $78.9\pm8.0$&   $25.4\pm3.2$&   $7.9\pm0.7$&   $21.1\pm2.1$&   $1.10$&   N \\
   M82&               3.4&           0.00094&              10.0&   $0.4\pm0.1$ &   $0.5\pm0.1$&   $4.0\pm0.3$&   $3.8\pm0.3$&   $352.8\pm53.5$&   $140.5\pm9.9$&   $537.2\pm41.2$&   $2.08$&   N \\
   MCG+12-02-001&              68.0&           0.01570&         $-$&   $2.6\pm0.5$ &   $4.1\pm1.1$&   $49.0\pm9.8$&   $52.9\pm8.0$&   $6.7\pm1.8$&   $4.4\pm0.9$&   $19.0\pm2.9$&   $1.53$&   N \\
   Mrk331&              80.3&           0.01848&              10.7&   $3.0\pm0.6$ &   $7.3\pm1.6$&   $31.1\pm5.4$&   $22.3\pm3.4$&   $8.6\pm1.9$&   $2.0\pm0.3$&   $5.8\pm0.9$&   $1$&   N \\
   NGC0253&               3.4&           0.00081&              10.4&   $0.3\pm0.1$ &   $0.9\pm0.1$&   $5.5\pm0.4$&   $5.1\pm0.4$&   $570.8\pm76.2$&   $191.6\pm13.3$&   $711.1\pm52.8$&   $2.19$&   Y \\
   NGC1068&              16.1&           0.00379&              11.0&   $1.3\pm0.3$ &   $5.6\pm0.4$&   $36.2\pm4.7$&   $33.8\pm3.2$&   $161.0\pm11.8$&   $57.3\pm7.5$&   $214.1\pm20.3$&   $1.87$&   Y \\
   NGC1365-NE&              23.1&           0.00546&         $-$&   $1.4\pm0.3$ &   $5.3\pm0.6$&   $48.0\pm1.2$&   $24.1\pm3.9$&   $73.8\pm8.6$&   $37.0\pm0.9$&   $74.1\pm12.0$&   $1.76$&   Y \\
   NGC1365-SW&              23.1&           0.00546&         $-$&   $1.4\pm0.3$ &   $7.3\pm0.7$&   $74.3\pm1.7$&   $29.6\pm4.8$&   $103.1\pm10.5$&   $57.3\pm1.3$&   $91.2\pm14.8$&   $2.79$&   Y \\
   NGC3256&              40.4&           0.00935&              10.8&   $4.2\pm0.8$ &   $9.2\pm1.1$&   $56.6\pm5.4$&   $128.9\pm13.3$&   $42.7\pm4.9$&   $14.3\pm1.4$&   $130.7\pm13.5$&   $1.29$&   N \\
   NGC5135&              59.3&           0.01369&              10.9&   $1.8\pm0.4$ &   $9.3\pm1.1$&   $40.6\pm3.7$&   $-$&   $20.0\pm2.3$&   $4.8\pm0.4$&   $-$&   $1.11$&   Y \\
   NGC6240&             106.8&           0.02448&              11.4&   $6.8\pm1.4$ &   $25.3\pm5.0$&   $79.3\pm8.1$&   $149.1\pm29.9$&   $16.9\pm3.4$&   $2.9\pm0.3$&   $21.9\pm4.4$&   $1$&   Y \\
   NGC7469&              70.8&           0.01632&              11.1&   $3.5\pm0.7$ &   $10.5\pm1.8$&   $41.4\pm3.4$&   $101.3\pm15.6$&   $15.9\pm2.7$&   $3.4\pm0.3$&   $33.6\pm5.2$&   $1.22$&   Y \\
   NGC7552&              22.8&           0.00536&              10.5&   $1.0\pm0.2$ &   $2.6\pm0.4$&   $13.0\pm0.9$&   $17.2\pm1.7$&   $37.8\pm6.3$&   $10.3\pm0.8$&   $54.5\pm5.5$&   $1.45$&   N \\
   NGC7582&              22.3&           0.00525&              10.6&   $0.7\pm0.1$ &   $2.7\pm0.4$&   $14.7\pm1.1$&   $21.6\pm2.3$&   $41.2\pm6.1$&   $12.2\pm0.9$&   $71.4\pm7.6$&   $1.87$&   Y \\
   NGC7771&              61.8&           0.01427&              11.2&   $2.4\pm0.5$ &   $8.5\pm1.1$&   $67.9\pm5.3$&   $77.9\pm11.7$&   $16.8\pm2.1$&   $7.4\pm0.6$&   $33.9\pm5.1$&   $1.68$&   N \\
   VV340A&             147.9&           0.03367&              11.2&   $5.3\pm1.1$ &   $33.6\pm8.9$&   $108.6\pm17.4$&   $-$&   $11.8\pm3.1$&   $2.1\pm0.3$&   $-$&   $1.08$&   N \\
   CenA&               7.7&           0.00183&              11.1&   $0.4\pm0.1$ &   $3.1\pm0.2$&   $21.0\pm2.1$&   $10.6\pm1.5$&   $392.1\pm26.2$&   $144.5\pm14.8$&   $291.1\pm41.2$&   $9.80$&   Y \\
   IC1623&              87.3&           0.02007&         $-$&   $4.7\pm0.9$ &   $15.4\pm2.7$&   $52.6\pm5.8$&   $77.6\pm11.8$&   $15.4\pm2.7$&   $2.9\pm0.3$&   $17.0\pm2.6$&   $1$&   N \\
   NGC0034&              85.3&           0.01962&              10.7&   $3.2\pm0.6$ &   $5.1\pm1.5$&   $24.9\pm2.4$&   $13.8\pm2.8$&   $5.4\pm1.5$&   $1.4\pm0.1$&   $3.2\pm0.7$&   $1.07$&   Y \\
   NGC0891-North&               7.4&           0.00176&         $-$&   $0.14\pm0.03$ &   $0.8\pm0.2$&   $6.4\pm1.0$&   $6.9\pm1.0$&   $107.5\pm20.7$&   $47.9\pm7.3$&   $207.8\pm31.2$&   $4.88$&   N \\
   NGC0891-South&               7.4&           0.00176&         $-$&   $0.13\pm0.03$ &   $1.7\pm0.3$&   $-$&   $-$&   $230.4\pm42.3$&   $-$&   $-$&   $-$&   N \\
   NGC2146-NUC&              12.6&           0.00298&         $-$&   $0.6\pm0.1$ &   $0.9\pm0.1$&   $7.1\pm0.2$&   $7.7\pm0.7$&   $42.9\pm6.8$&   $18.4\pm0.6$&   $79.3\pm7.2$&   $1.65$&   N \\
   NGC2146-NW&              12.6&           0.00298&         $-$&   $0.6\pm0.1$ &   $1.2\pm0.2$&   $7.5\pm0.2$&   $8.9\pm0.8$&   $55.8\pm7.1$&   $19.4\pm0.6$&   $91.4\pm8.4$&   $1.69$&   N \\
   NGC2146-SE&              12.6&           0.00298&         $-$&   $0.6\pm0.1$ &   $1.1\pm0.1$&   $7.8\pm0.2$&   $8.1\pm1.6$&   $49.4\pm6.9$&   $20.2\pm0.6$&   $83.8\pm16.8$&   $1.76$&   N \\
   NGC3227&              16.3&           0.00386&              10.2&   $0.07\pm0.01$ &   $0.9\pm0.2$&   $5.8\pm0.6$&   $-$&   $25.0\pm6.7$&   $9.0\pm1.0$&   $-$&   $2.27$&   Y \\
   \enddata
   \tablecomments{\footnotesize 1 - ID; 2 - Distance; 3 - Redshift; 4 - Stellar mass from 2MASS $K_{\rm s}$-band photometry (\citealt{chabrier_2003} IMF, uncertainty $\sim0.2$~dex); 5~-~Total infrared luminosity integrated within $8-1000$~$\mu$m: $L_{\rm IR} = L_{\rm FIR}(40-400\,\mu\mathrm{m})\times1.2$; 6 - 8: \lprimeci, \lprimecoone, \lprimecotwo; 9 - Galaxy integrated \cione\ velocity integrated flux from FTS observations \citep{liu_2015}; 10, 11 - Galaxy integrated \coone\ and \cotwo\ fluxes from ground-based facilities. These values are signal-to-noise weighted means of all the available measurements in  a fixed beam of 43.5'' as in Table 3 of \cite{kamenetzky_2016}, corrected by the beam factor in column (12); 12 - Beam correction for low-$J$ \co\ fluxes in columns (10, 11): $\eta_{\rm \,beam} = I_{\rm{[C\,\scriptscriptstyle{I}\scriptstyle{]}}}(\mathrm{L15,
 total})/I_{\rm{[C\,\scriptscriptstyle{I}\scriptstyle{]}}}(\mathrm{K16,
 43.5" beam})$; 13 - Active galactic nuclei entry in
\cite{veron-cetty_2010}: Y(es)/N(o).\\
\\
\\
\textbf{References:} Cols 5, 9: \cite{liu_2015}, this work; Cols 10,
11: \cite{kamenetzky_2016}; Col 13: \cite{veron-cetty_2010}.}
 \end{deluxetable}

 \begin{deluxetable}{lccc}
   \tabletypesize{\normalsize}
   \tablecolumns{4}
   \tablecaption{\normalsize  Local galaxies (continue).\label{tab:mgaslocal}}
   \smallskip
   \tablehead{
     \colhead{ID} & 
     \colhead{\mci}&
     \colhead{\mgas(CO)}&
     \colhead{\ciabundance$_{\rm(CO)}$}\\
     \colhead{}&
     \colhead{$10^{5}$~\msun}& 	
     \colhead{$10^{9}$~\msun}&
     \colhead{$\times10^{-5}$}\\
     \colhead{(1)}&
     \colhead{(2)}& 	
     \colhead{(3)}&
     \colhead{(4)}\\
   }
   \startdata
   Arp193&   $13.0\pm4.1$&   $3.2\pm1.5$&   $9.3\pm5.3$ \\
   Arp220&   $17.3\pm3.5$&   $6.5\pm3.0$&   $6.0\pm3.1$ \\
   Arp299-A&   $6.0\pm1.5$&   $2.3\pm1.1$&   $6.0\pm3.2$ \\
   Arp299-B&   $9.1\pm2.7$&   $1.5\pm0.7$&   $13.7\pm7.7$ \\
   Arp299-C&   $8.4\pm2.0$&   $1.4\pm0.7$&   $13.5\pm7.2$ \\
   CGCG049-057&   $2.9\pm0.8$&   $-$&   $-$ \\
   ESO173-G015&   $8.8\pm1.9$&   $-$&   $-$ \\
   IRASF18293-3413&   $26.0\pm3.3$&   $9.5\pm4.5$&   $6.2\pm3.0$ \\
   M82&   $0.7\pm0.1$&   $0.3\pm0.1$&   $4.9\pm2.4$ \\
   MCG+12-02-001&   $5.1\pm1.3$&   $3.9\pm2.0$&   $3.0\pm1.7$ \\
   Mrk331&   $9.2\pm2.0$&   $2.5\pm1.2$&   $8.4\pm4.5$ \\
   NGC0253&   $1.1\pm0.1$&   $0.4\pm0.2$&   $5.8\pm2.8$ \\
   NGC1068&   $7.0\pm0.5$&   $2.9\pm1.4$&   $5.4\pm2.6$ \\
   NGC1365-NE&   $6.6\pm0.8$&   $3.8\pm1.8$&   $3.9\pm1.8$ \\
   NGC1365-SW&   $9.2\pm0.9$&   $5.9\pm2.7$&   $3.5\pm1.7$ \\
   NGC3256&   $11.5\pm1.3$&   $4.5\pm2.1$&   $5.8\pm2.8$ \\
   NGC5135&   $11.6\pm1.3$&   $3.2\pm1.5$&   $8.1\pm3.9$ \\
   NGC6240&   $31.6\pm6.3$&   $6.3\pm3.0$&   $11.3\pm5.8$ \\
   NGC7469&   $13.1\pm2.3$&   $3.3\pm1.5$&   $9.0\pm4.5$ \\
   NGC7552&   $3.3\pm0.5$&   $1.0\pm0.5$&   $7.1\pm3.5$ \\
   NGC7582&   $3.4\pm0.5$&   $1.2\pm0.6$&   $6.6\pm3.2$ \\
   NGC7771&   $10.6\pm1.3$&   $5.4\pm2.5$&   $4.4\pm2.1$ \\
   VV340A&   $42.0\pm11.1$&   $8.7\pm4.2$&   $10.9\pm6.1$ \\
   CenA&   $3.9\pm0.3$&   $1.7\pm0.8$&   $5.3\pm2.5$ \\
   IC1623&   $19.2\pm3.3$&   $4.2\pm2.0$&   $10.4\pm5.2$ \\
   NGC0034&   $6.4\pm1.8$&   $2.0\pm0.9$&   $7.3\pm4.0$ \\
   NGC0891-North&   $1.0\pm0.2$&   $0.5\pm0.2$&   $4.4\pm2.3$ \\
   NGC0891-South&   $2.1\pm0.4$&   $-$&   $-$ \\
   NGC2146-NUC&   $1.1\pm0.2$&   $0.6\pm0.3$&   $4.5\pm2.2$ \\
   NGC2146-NW&   $1.5\pm0.2$&   $0.6\pm0.3$&   $5.6\pm2.7$ \\
   NGC2146-SE&   $1.3\pm0.2$&   $0.6\pm0.3$&   $4.7\pm2.3$ \\
   NGC3227&   $1.1\pm0.3$&   $0.5\pm0.2$&   $5.4\pm2.9$ \\
   \enddata
   \tablecomments{\footnotesize 1 - ID; 2 - Mass of \ci\ (from
     \lprimeci\ assuming $T_{\rm ex} = 30$~K); 3 - Gas mass from
     \coone, assuming $\alpha_{CO} = 0.8$~\msun/(\kkmspc) (with an
     uncertainty of 0.2~dex); 4 - Atomic carbon abundance adopting \mgas(\co) and removing the $1.36\times$ contribution
  of helium.}
 \end{deluxetable}

\begin{deluxetable}{lccccccccccccccc}
   \tabletypesize{\tiny}
   \tablewidth{4pt}
   \tablecolumns{15}
   \tablecaption{Sub-millimeter galaxies and QSOs at $z\gtrsim2.5$. \label{tab:smgs}}
   \smallskip
   \tablehead{\\
     \colhead{ID} & 
     \colhead{$z_{\rm spec}$}&
     \colhead{$\mu_{\rm grav}$}&
     \colhead{\lir}&
     \colhead{\lfir}&
     \colhead{\lprimeci}&
     \colhead{$L'_{\rm CO(3-2)}$}&
     \colhead{\ici}&
     \colhead{$I_{\rm CO(3-2)}$}&
     \colhead{\mdust}&
     \colhead{\mci}&
     \colhead{$\displaystyle \frac{\rm[ C\,\scriptstyle
           I]}{[\mathrm{H}_2]}_{\rm dust}$}&
     \colhead{$\displaystyle \frac{\rm[ C\,\scriptstyle
         I]}{[\mathrm{H}_2]}_{\rm CO}$}&
     \colhead{$f_{\rm AGN}$}&
     \colhead{QSOs} \tiny\\
     \colhead{}&
     \colhead{}&
     \colhead{}&
     \colhead{$10^{12}\,L_{\odot}$}&
     \colhead{$10^{12}\,L_{\odot}$}&
     \colhead{$10^{10}\,\mathrm{K\,km\,s^{-1}\,pc^2}$}&
     \colhead{$10^{10}\,\mathrm{K\,km\,s^{-1}\,pc^2}$}&
     \colhead{$\mathrm{Jy\,km\,s^{-1}}$}&
     \colhead{$\mathrm{Jy\,km\,s^{-1}}$}&
     \colhead{$10^{8}\,M_\odot$}&
     \colhead{$10^{6}\,M_\odot$}&
     \colhead{$\times10^{-5}$}&
     \colhead{$\times10^{-5}$}&
     \colhead{}&
     \colhead{}\\
     \colhead{(1)}&
     \colhead{(2)}& 	
     \colhead{(3)}&
     \colhead{(4)}&
     \colhead{(5)}&
     \colhead{(6)}&
     \colhead{(7)}&
     \colhead{(8)}&
     \colhead{(9)}&
     \colhead{(10)}&
     \colhead{(11)}&
     \colhead{(12)}&
     \colhead{(13)}&
     \colhead{(14)}&
     \colhead{(15)}\\
   }
   \rotate
   \startdata 
   SMMJ02399-0136&           2.808&               2.5&    $9.2\pm0.9$ &  $6.1\pm0.6$ &   $1.48\pm0.16$&   $4.88\pm0.63$&   $1.9\pm0.2$&   $3.1\pm0.4$&   $32.3\pm1.1$&   $18.4\pm1.9$&   $4.3\pm2.0$&   $5.6\pm2.7$&              0.09&    N \\
   APM08279+5255&           3.911&              80&    $16.0\pm0.4$ &   $3.6\pm0.1$ &   $0.04\pm0.01$&   $-$&   $0.9\pm0.1$&   $-$&   $-$&   $0.5\pm0.1$&  $-$&   $-$&              0.99&    Y \\
   RXJ0911+0551&           2.796&              20&    $10.6\pm1.6$ &       $2.5\pm0.4$ &   $0.20\pm0.03$&   $0.39\pm0.06$&   $2.1\pm0.3$&   $2.0\pm0.3$&   $4.7\pm0.7$&   $2.5\pm0.4$&   $4.0\pm2.0$&   $9.5\pm4.8$&              0.98&    Y \\
   F10214&           2.285&              10&    $33.0\pm1.5$ &                   $10.1\pm0.5$ & $0.27\pm0.05$&   $1.16\pm0.22$&   $2.0\pm0.4$&   $4.2\pm0.8$&   $6.6\pm0.5$&   $3.4\pm0.7$&   $3.9\pm2.0$&   $4.3\pm2.3$&              0.74&    Y \\
   SMMJ123549+6215&           2.202&               1&    $4.2\pm1.3$ &   $3.6\pm1.1$ &   $1.41\pm0.26$&   $4.15\pm0.52$&   $1.1\pm0.2$&   $1.6\pm0.2$&   $64.6\pm20.1$&   $17.6\pm3.2$&   $2.1\pm1.2$&   $6.2\pm3.2$&              $-$&    N \\
   BRI1335-0417&           4.407&               1&    $49.6\pm3.8$ &         $20.8\pm1.6$ & $<8.83$&   $-$&   $<2.2$&   $-$&   $59.8\pm4.6$&   $<110.2$&   $<13.9$&   $-$&              0.29&    Y \\
   SMMJ14011+0252&           2.565&               4$^{\dagger}$&    $3.7\pm0.1$ & $2.9\pm0.1$ &   $0.75\pm0.12$&   $2.36\pm0.25$&   $1.8\pm0.3$&   $2.8\pm0.3$&   $7.12\pm0.02$&   $9.4\pm1.6$&  $9.9\pm4.9$&   $5.8\pm2.9$&              0.05&    N \\
   Cloverleaf&           2.558&              11&    $26.9\pm0.6$ &               $6.3\pm0.1$ &   $0.59\pm0.09$&   $4.03\pm0.06$&   $3.9\pm0.6$&   $13.2\pm0.2$&   $8.7\pm0.4$&   $7.3\pm1.1$&   $6.4\pm3.1$&   $2.7\pm1.3$&              0.98&    Y \\
   SMMJ16359+6612&           2.517&              $22\pm2$&    $0.74\pm0.03$ &   $0.55\pm0.02$ & $0.12\pm0.02$&   $0.42\pm0.03$&   $1.7\pm0.3$&   $2.8\pm0.2$&   $4.60\pm0.04$&   $1.6\pm0.3$&   $2.6\pm1.3$&   $5.5\pm2.7$&             0.12&    N \\
   SMMJ163650+4057&           2.385&               1&    $8.4\pm0.4$ &   $5.2\pm0.2$ &   $<1.06$&   $4.77\pm0.60$&   $<0.7$&   $1.6\pm0.2$&   $38.4\pm1.8$&   $<13.2$&   $<2.6$&   $<4.1$&              $-$&    N \\
   SMMJ163658+4105&           2.452&               1&    $9.5\pm0.7$ &   $8.0\pm0.6$ &   $1.39\pm0.31$&   $5.62\pm0.62$&   $0.9\pm0.2$&   $1.8\pm0.2$&   $22.9\pm1.7$&   $17.3\pm3.9$&   $5.7\pm2.9$&   $4.5\pm2.4$&              $-$&    N \\
   MM18423+5938&           3.930&              20&    $10.2\pm1.0$ &     $5.0\pm0.5$ &   $0.39\pm0.08$&   $-$&   $2.3\pm0.5$&   $-$&   $7.88\pm0.04$&   $4.8\pm1.0$&   $4.6\pm2.4$&   $-$&              0.18&    N \\
   SMMJ213511-0102&           2.326&              $32.5\pm4.5$&    $2.4\pm0.1$ &   $2.1\pm0.1$ & $0.69\pm0.02$&   $1.16\pm0.01$&   $16.0\pm0.5$&   $13.2\pm0.1$&   $8.7\pm0.2$&   $8.7\pm0.3$&   $7.5\pm3.5$&   $11.0\pm5.1$&              $-$&    N \\
   PSSJ2322+1944&           4.120&               $5.3\pm0.3$&    $17.8\pm1.7$ &   $5.3\pm0.5$ & $0.55\pm0.08$&   $-$&   $0.8\pm0.1$&   $-$&   $14.5\pm1.4$&   $6.8\pm1.0$&   $3.6\pm1.8$&   $-$&              0.88&    Y \\
   ID141&           4.243&              20$^{\dagger}$&    $5.4\pm0.3$ &      $4.3\pm0.2$ &   $0.53\pm0.17$&   $-$&   $2.8\pm0.9$&   $-$&   $7.9\pm0.4$&   $6.6\pm2.1$&   $6.3\pm3.5$&   $-$&              $-$&    N \\
   SXDF7&           2.529&               1&    $8.1\pm0.4$ &                       $6.0\pm0.3$ &    $1.30\pm0.33$&   $-$&   $0.8\pm0.2$&   $-$&   $15.2\pm0.7$&   $16.2\pm4.1$&   $8.1\pm4.3$&   $-$&              0.01&    N \\
   SXDF11&           2.282&               1&    $3.3\pm0.2$ &                     $2.4\pm0.1$ &    $<0.95$&   $-$&   $<0.7$&   $-$&   $23.9\pm1.4$&   $<11.9$&   $<3.8$&   $-$&              $-$&    N \\
   SXDF4a$^{\ddagger}$&           2.030&               1&    $4.6\pm0.2$ &   $3.3\pm0.2$ &   $0.78\pm0.22$&   $-$&   $0.7\pm0.2$&   $-$&   $21.5\pm1.1$&   $9.7\pm2.8$&   $3.4\pm1.9$&   $-$&              $-$&    N \\
   SXDF4b$^{\ddagger}$&           2.027&               1&    $4.6\pm0.2$ &   $3.3\pm0.2$ &   $<0.77$&   $-$&   $<0.7$&   $-$&   $21.5\pm1.1$&   $<9.7$&   $<3.4$&   $-$&              $-$&    N \\
   SA22.96&           2.517&               1&    $7.6\pm0.3$ &                    $6.6\pm0.2$ &    $1.13\pm0.32$&   $5.23\pm1.63$&   $0.7\pm0.2$&   $1.6\pm0.5$&   $26.9\pm1.0$&   $14.1\pm4.0$&   $4.0\pm2.2$&   $4.0\pm2.5$&              $-$&    N \\
   \enddata
   \tablecomments{\footnotesize 1 - ID; 2 - Spectroscopic redshift; 3
     -  Gravitational magnification factor; 4 - Total infrared
     luminosity integrated within $8-1000$~$\mu$m. This value is due to star
     formation only (i.e., corrected for torus emission) for SMGs (`N'
     in column 15), while it represents the SF+AGN emission for QSOs
     (`Y' in column 15); 5 - Far-infrared luminosity integrated within
     $40-400$~$\mu$m. This value is due to star
     formation only (i.e., corrected for torus emission) for SMGs (`N'
     in column 15), while it represents the SF+AGN emission for QSOs
     (`Y' in column 15);
    6, 7 - Galaxy-integrated \lprimeci\ and
     $L'_{\rm CO(3-2)}$; 8, 9 - Velocity-integrated \cione\ and
     \cothree\ fluxes; 10 - Dust mass; 11 - Mass of \ci\ (from
     \lprimeci\ assuming $T_{\rm ex} = 30$~K); 12 - Atomic carbon
     abundances computed assuming \mgas(dust, $\delta_{\rm GDR} =
     30$)  and removing the $1.36\times$ contribution
  of helium; 13 - Atomic carbon
     abundances computed assuming \mgas(CO, $\alpha_{\rm
       CO}=0.8$~\msun/(\kkmspc), $r_{31} = 0.52$)  and removing the $1.36\times$ contribution
  of helium; 14 - Fraction infrared AGN emission: $L_{\rm
       AGN+SF} = L_{\rm IR} / (1-f_{\rm AGN})$; 15 - Quasar activity:
     Y(es)/N(o).\\
     \\
     \\
     Upper limits are at $3\sigma$.\\ 
     All values have been corrected for gravitational magnification.\\
     $^\dagger$: From \citep{alaghband-zadeh_2013}.\\
     $^{\ddagger}$: This source is unresolved at $850$~$\mu$m, but two
     components are evident in the H${\alpha}$ and \cofour\ maps. The
     photometry refers to the unresolved source.\\
     \\
     \textbf{References:} Cols 1-3, 8, 9, 15: \citet[W11]{walter_2011},
     \citet[AZ13]{alaghband-zadeh_2013}, and references therein; Cols
     4, 5
     (photometry): SMMJ02399-0136, SMMJ14011+0252, SMMJ16359+6612: \cite{magnelli_2012};
     APM08279+5255, RXJ0911+0551, F10214, Cloverleaf, PSSJ2322+1944: \cite{stacey_2018};
     SMMJ123549+6215:
     W11, \cite{kirkpatrick_2012}, \cite{ivison_2011}; 
     BRI1335-0417: \cite{guilloteau_1997},
     \cite{carilli_1999}, W11, \cite{wagg_2014}; SMMJ163650+4057, SMMJ163658+4105:
     \cite{kovacs_2006}, \cite{tacconi_2006}, \cite{efstathiou_2009},
     W11; MM18423+5938: \cite{lestrade_2010}, \cite{catalano_2014};
     \cite{mckean_2011}; SMMJ213511-0102: \cite{ivison_2010b},
     \cite{swinbank_2010}; ID141: \cite{cox_2011}; SXDF7, SXDF11, SXDF4a+b:
     \cite{ivison_2007},
     \cite{alaghband-zadeh_2012,alaghband-zadeh_2013}; SA22.96:
     \cite{menendez-delmestre_2009}, \cite{alaghband-zadeh_2012,alaghband-zadeh_2013}.}
 \end{deluxetable}

 \begin{deluxetable}{lccccccccccccccc}
   \tabletypesize{\tiny}
   \tablecolumns{15}
   \tablecaption{SPT sub-millimeter galaxies at $z\sim4$. \label{tab:sptsmgs}}
   \tablehead{\\
     \colhead{ID} & 
     \colhead{$z_{\rm spec}$}&
     \colhead{$\mu_{\rm grav}$}&
     \colhead{$M_\star$}&     
     \colhead{\lir}&
     \colhead{\lfir}&
     \colhead{\lprimeci}&
     \colhead{\lprimecotwo}&
     \colhead{\ici}&
     \colhead{$I_{\rm CO(2-1)}$}&
     \colhead{\mdust}&
     \colhead{\mci}&
     \colhead{$\displaystyle \frac{\rm[ C\,\scriptstyle
           I]}{[\mathrm{H}_2]}_{\rm dust}$}&
     \colhead{$\displaystyle \frac{\rm[ C\,\scriptstyle
         I]}{[\mathrm{H}_2]}_{\rm CO}$}&
     \colhead{$f_{\rm AGN}$} \tiny\\
     \colhead{}&
     \colhead{}&
     \colhead{}&
     \colhead{$\scriptstyle 10^{10}\,M_{\odot}$}&
     \colhead{$\scriptstyle 10^{12}\,L_{\odot}$}&
     \colhead{$\scriptstyle 10^{12}\,L_{\odot}$}&
     \colhead{$\scriptstyle 10^{10}\,\mathrm{K\,km\,s^{-1}\,pc^2}$}&
     \colhead{$\scriptstyle 10^{10}\,\mathrm{K\,km\,s^{-1}\,pc^2}$}&
     \colhead{$\scriptstyle \mathrm{Jy\,km\,s^{-1}}$}&
     \colhead{$\scriptstyle \mathrm{Jy\,km\,s^{-1}}$}&
     \colhead{$\scriptstyle 10^{8}$~\msun}&
     \colhead{$\scriptstyle 10^{6}$~\msun}&
     \colhead{$\scriptstyle \times10^{-5}$}&
     \colhead{$\scriptstyle \times10^{-5}$}&
     \colhead{}\\
     \colhead{(1)}&
     \colhead{(2)}& 	
     \colhead{(3)}&
     \colhead{(4)}&
     \colhead{(5)}&
     \colhead{(6)}&
     \colhead{(7)}&
     \colhead{(8)}&
     \colhead{(9)}&
     \colhead{(10)}&
     \colhead{(11)}&
     \colhead{(12)}&
     \colhead{(13)}&
     \colhead{(14)}&
     \colhead{(15)}\\
   }
   \rotate
   \startdata  
SPT0113-46&           4.233&              $23.9\pm0.5$&            $-$&    $1.5\pm0.1$ &   $1.2\pm0.1$ &  $0.53\pm0.11$&   $1.22\pm0.09$&   $3.4\pm0.7$&   $1.7\pm0.1$&   $11.6\pm0.1$&   $6.6\pm1.3$&   $4.3\pm2.2$&   $12.9\pm6.6$&             $-$ \\
SPT0125-50&           3.959&              $14.1\pm0.5$&            $-$&    $10.4\pm0.5$ & $4.9\pm0.2$ &  $0.57\pm0.13$&   $-$&   $2.4\pm0.5$&   $-$&   $19.0\pm2.0$&   $7.1\pm1.6$&   $2.8\pm1.5$&   $-$&              0.16 \\
SPT0300-46&           3.596&               $5.7\pm0.4$&            $-$&    $8.8\pm0.2$ &    $6.9\pm0.2$ &  $<1.21$&   $-$&   $1.8\pm0.8$$^\dagger$&   $-$&   $27.3\pm2.0$&   $<15.1$&   $<4.2$&   $-$&              $-$ \\
SPT0345-47&           4.296&               $8.0\pm0.5$&            $-$&    $36.0\pm1.7$ &  $15.3\pm0.7$& $<0.50$&   $3.95\pm0.44$&   $<1.0$&   $1.8\pm0.2$&   $17.0\pm0.9$&   $<6.2$&   $<2.7$&   $<3.7$&              $-$ \\
SPT0418-47&           4.225&              $32.7\pm2.7$&            $-$&    $3.0\pm0.1$ &   $2.4\pm0.1$ &  $0.28\pm0.07$&   $0.68\pm0.06$&   $2.5\pm0.6$&   $1.3\pm0.1$&   $4.4\pm0.1$&   $3.5\pm0.9$&   $6.1\pm3.2$&   $12.3\pm6.6$&              $-$ \\
SPT0441-46&           4.477&              $12.7\pm1.0$&            $-$&    $7.2\pm0.6$ &   $3.9\pm0.4$ &  $<0.72$&   $1.40\pm0.21$&   $1.8\pm0.7$$^\dagger$&   $0.9\pm0.1$&   $14.7\pm0.7$&   $<9.0$&   $<4.6$&   $<15.2$&              $-$ \\
SPT0459-59&           4.799&               $3.6\pm0.3$ &            $-$&    $18.0\pm1.0$ & $9.8\pm0.5$ &  $3.09\pm0.89$&   $6.37\pm0.46$&   $2.4\pm0.7$&   $1.1\pm0.1$&   $52.8\pm2.1$&   $38.5\pm11.1$&   $5.5\pm3.0$&   $14.4\pm7.9$&              $-$ \\
SPT0529-54&           3.369&              $13.2\pm0.5$&            $-$&    $3.5\pm0.1$ &   $3.0\pm0.1$ &  $0.57\pm0.11$&   $-$&   $2.9\pm0.5$&   $-$&   $33.3\pm0.8$&   $7.1\pm1.3$&   $1.6\pm0.8$&   $-$&              $-$ \\
SPT0532-50&           3.399&              $10.0\pm0.6$&            $-$&    $7.6\pm0.7$ &   $5.7\pm0.5$ &  $0.85\pm0.20$&   $-$&   $3.2\pm0.8$&   $-$&   $57.4\pm1.3$&   $10.6\pm2.5$&   $1.4\pm0.7$&   $-$&              0.62 \\
SPT2103-60&           4.436&              $27.8\pm1.8$&            $-$&    $1.80\pm0.03$ &   $1.51\pm0.02$ &  $0.45\pm0.11$&   $1.06\pm0.17$&   $3.1\pm0.8$&   $1.6\pm0.2$&   $6.4\pm0.1$&   $5.6\pm1.4$&   $6.6\pm3.4$&   $12.5\pm6.8$&              $-$ \\
SPT2132-58&           4.768&               $5.7\pm0.5$&            $-$&    $16.8\pm0.4$ &  $7.7\pm0.2$ &  $<0.69$&   $3.08\pm0.25$&   $0.8\pm0.3$$^\dagger$&   $0.8\pm0.1$&   $32.0\pm0.8$&   $<8.6$&   $<2.0$&   $<6.7$&              $-$ \\
SPT2146-55&           4.567&               $6.7\pm0.4$&             $8.0^{+19.0}_{-6.0}$&    $11.4\pm0.4$ &   $5.9\pm0.2$ & $1.73\pm0.45$&   $2.74\pm0.46$&   $2.7\pm0.7$&   $0.9\pm0.2$&   $20.9\pm1.5$&   $21.6\pm5.6$&   $7.8\pm4.2$&   $18.7\pm10.4$&               0.13 \\
SPT2147-50&           3.760&               $6.6\pm0.4$&            $2.0^{+1.8}_{-0.9}$&    $7.5\pm1.1$ &       $5.9\pm0.9$ & $0.95\pm0.28$&   $2.70\pm0.54$&   $2.0\pm0.6$&   $1.2\pm0.2$&   $19.5\pm0.5$&   $11.9\pm3.5$&   $4.6\pm2.5$&   $10.5\pm6.1$&              $-$ \\
   \enddata
   \tablecomments{\footnotesize 1 - ID; 2 - Spectroscopic redshift; 3
     -  Gravitational magnification factor; 4 - Stellar mass from
     \cite{ma_2015}. We did not correct it for the minimal differences
     in the choice of IMF (\cite{chabrier_2003} vs \cite{kroupa_2001}) and stellar population
     models (\cite{bruzual_2003} vs \cite{maraston_2005}, see 
     Section 6.1 in \citealt{ma_2015}); 5 - Total infrared luminosity integrated within
     $8-1000$~$\mu$m due to star formation (i.e., corrected for torus
     emission); 6 - Far-infrared luminosity integrated within
     $40-400$~$\mu$m due to star formation (i.e., corrected for torus
     emission); 7, 8 - Galaxy-integrated \lprimeci\ and
     $L'_{\rm CO(2-1)}$; 9 - Velocity-integrated \cione\ fluxes;
     10 - Velocity-integrated \cotwo\ fluxes; 11 - Dust mass; 12 - Mass of \ci\ (from
     \lprimeci\ assuming $T_{\rm ex} = 30$~K); 13 - Atomic carbon
     abundances computed assuming \mgas(dust, $\delta_{\rm GDR} =
     30$)  and removing the $1.36\times$ contribution
  of helium; 14 - Atomic carbon
     abundances computed assuming \mgas(CO, $\alpha_{\rm
       CO}=0.8$~\msun/(\kkmspc), $r_{21} = 0.84$)  and removing the $1.36\times$ contribution
  of helium; 15 - Fraction infrared AGN emission: $L_{\rm
       AGN+SF} = L_{\rm IR} / (1-f_{\rm AGN})$.\\
     \\ 
     \\
     Upper limits are at $3\sigma$.\\ 
     All values have been corrected for gravitational magnification.\\
     $^\dagger$: We report the nominal flux measurements presented in
     \cite{bothwell_2017}, but we adopt the $3\sigma$ upper limits for
     the analysis.\\
   \\
   \textbf{References:} Cols 1-3, 9: \cite{bothwell_2017}; Col 4:
   \cite{ma_2015}; Cols 5, 6 (photometry): \cite{weiss_2013}; Col 10: \cite{aravena_2016}.}
 \end{deluxetable}

\end{document}